\documentclass[11pt, a4paper]{article}

\pdfoutput=1
\RequirePackage[numbers]{natbib}

\usepackage[bookmarks=false]{hyperref}
\usepackage{amsthm,amsfonts,amssymb,amsmath}
\usepackage{mathrsfs}

\usepackage{tikz}
\usepackage{tikz-3dplot}

\usepackage{multirow}

\usepackage{listings}
\usepackage[latin1]{inputenc}
\usepackage[T1]{fontenc}
\usepackage[english]{babel}

\usepackage{subcaption}

\usepackage{articlestyle}

\usepackage{algorithm}
\usepackage{authblk}

\usepackage{fullpage}

\usepackage{ifpdf}
\usepackage{units}

\usepackage{mathtools}
\mathtoolsset{showonlyrefs}

\usepackage{xr}

\usepackage[normalem]{ulem}

\title{Joint spatial modeling of significant wave height and wave period using the SPDE approach}
\author[1]{Anders Hildeman}
\author[1]{David Bolin}
\author[1]{Igor Rychlik}
\date{}

\affil[1]{Department of Mathematical Sciences, Chalmers University of Technology and University of Gothenburg, Sweden}

\begin{document}

\maketitle

\setcounter{tocdepth}{2}

\begin{abstract}
\label{sec:abstract}
The ocean wave distribution in a specific region of space and time is described by
its sea state. 
Knowledge about the sea states a ship encounters on a journey can be used to assess various parameters of risk and wear associated with the journey. 
Two important characteristics of the sea state are the significant wave height and mean wave period. We propose a joint spatial model of these two quantities on the north Atlantic ocean. 
The model describes the distribution of the logarithm of the two quantities as a bivariate Gaussian random field. This random field is modeled as a solution to a system of coupled stochastic partial differential equations. The bivariate random field can model a wide variety of non-stationary anisotropy and allows for arbitrary, and different, differentiability for the two marginal fields.

The parameters of the model are estimated on data of the north Atlantic using a stepwise maximum likelihood method. 
The fitted model is used to derive the distribution of accumulated fatigue damage for a ship sailing a transatlantic route. 
Also, a method for estimating the risk of capsizing due to broaching-to, based on the joint distribution of the two sea state characteristics, is investigated. 
The risks are calculated for a transatlantic route between America and Europe using both data and the fitted model.

The results show that the model compares well with observed data. Also, it shows that the bivariate model is needed and cannot simply be approximated by a model of significant wave height alone.

\end{abstract}


\section{Introduction}
\label{sec:introduction}
The sea state characterizes the stochastic behavior of ocean waves in a region in space and time. 
Explicit knowledge of the sea state allows for quantitative assessments of profits, costs, and risks associated with naval logistics, fishing, marine operations, and other applications affected by the sea surface conditions.

Let us denote the spatio-temporal stochastic process of sea surface elevation as $W(\psp, t)$, where $\psp \in \gspace$, $t \in [0,\mathcal{T}]$. Here, $\gspace$ is a small region in space and $[0,\mathcal{T}]$ is a small interval in time, typically from 20 minutes up to about 3 hours.
The distribution of $W$ is equivalent to the sea state at $\gspace \times [0,\mathcal{T}]$.
In general, a spatio-temporal stochastic process can be very complex to model. 
However, for waves in deep water, the sea surface elevation is often approximated by means of Gaussian fields. Furthermore, if $\gspace$ and $\mathcal{T}$ are small enough, $W$ will be a stationary Gaussian process. For most applications, the quantities of interest are the deviations from the sea level, hence the mean value is of no interest. Then, $W$ could be modeled as a centered stationary Gaussian process and is completely characterized by the directional spectrum $S(\omega, \theta)$. Here $\omega \ge 0$ is the angular frequency of the waves and $\theta \in [0, 2\pi]$ is the direction \citep{lit:aberg}. 

In this paper we are concerned with applications related to ship safety. For such applications we are mainly interested in sea states where a dominant part of the wave energy is propagating in a narrow band of directions. Hence, we will make the approximation $S(\omega, \theta) = S(\omega)\delta(\theta-\theta_0)$, where $S(\omega) = \int_0^{2\pi}S(\omega, \theta)d\theta$ is the temporal spectrum, 
$\theta_0$ is the direction the waves are, approximately, propagating from and $\delta$ is the Dirac delta function. This approximation is known as a \textit{long crested sea}, for which the sea state is completely characterized by its temporal spectrum and a wave direction. 


For most applications, a few scalar valued quantities are enough to characterize $S$. For example, the popular parametric Bretschneider spectrum \citep{lit:bretschneider}, which has been shown to explain the important characteristics of sea states for a wide range of applications and spatial regions, is fully characterized by the \textit{significant wave height} $H_s$ and the \textit{peak wave period} $T_p$. The Bretschneider spectrum is defined as 
\begin{equation}\label{PM}
S(\omega) = c\omega^{-5}\exp \left( -1.25 \frac{\omega_p^4}{\omega^4} \right), \qquad 
c= \frac{1.25}{4} \,H_s^2\omega_p^4, \qquad \omega_p=2\pi\,/T_p.
\end{equation}
Here, $H_s = 4\sqrt{\Var[W(\psp, t)}]$ is four times the standard deviation of the sea surface elevation. It is a quantity summarizing the distribution of wave heights of apparent waves and is measured in units of length, in this paper in meters $[m]$. The significant wave height is in general the most important single quantity when assessing risks to ships in a given sea state. 
The peak wave period is defined as the wave period with the highest energy, 
\begin{align}
    T_p = \arg \max_{\omega > 0} \frac{2\pi}{S(\omega)},
\end{align}
and summarizes the distribution of wave periods of apparent waves and is measured in units of time, in our paper in seconds $[s]$.
Two other popular quantities summarizing the distribution of wave periods are the \textit{mean wave period}, $T_1$, and \textit{mean zero-crossing period}, $T_z$, defined as 
\begin{align}
    T_1 = 2\pi \frac{\int_{0}^{\infty}\omega^{-1}\,S(\omega)d\omega  }{\int_{0}^{\infty}S(\omega)d\omega  }\qquad
    T_z = 2\pi \sqrt{\frac{\int_{0}^{\infty}S(\omega)d\omega  }{\int_{0}^{\infty}\omega^2\,S(\omega)d\omega  }}.
\end{align}
In words, $T_1$ is the mean of the period spectrum while $T_z$ is the mean time between a zero upcrossing and the consecutive, for a fixed point in space.
Under the assumption of a Bretschneider spectrum, these three quantities are related as $T_p = 1.408 \cdot T_z = 1.2965 \cdot T_1$.
Since all three quantities are proportional to each other under the assumption of a Bretschneider spectrum, we will in this paper use the notation, $T$, to denote a quantity of the wave period without explicitly stating which. 
Hence, as long as the Bretschneider spectrum is a reasonable approximation, all information about the sea state is encoded in the two quantities $H_s$ and $T$.

The problem with using a the Bretschneider spectrum to model the sea state is that it assumes stationarity, which is not valid for large spatial regions. This is often solved by assuming that the parameters $H_s$ and $T$ are spatially varying. The main contribution of this work is to propose a joint spatial model for $H_s$ and $T$, which can be used to describe the sea states for large regions. 




Probabilistic models of $H_s$ and $T$ jointly for a fixed point in space and time have been studied extensively. \citet{lit:ochi} showed that a bivariate log-normal distribution fits the bulk of the marginal probability distributions of $H_s$ and $T$ for data from the north Atlantic. Other approaches are to use Placket-models \citep{lit:placket,lit:athanassoulis}, or more general Box-Cox transformations \citep{lit:box} and then model the transformed values with a bivariate Gaussian distribution. Conditional modeling approaches have also been proposed where $H_s$ is first modeled and $T$ is modeled conditional on $H_s$ \citep{lit:soares, lit:lucas, lit:vanem}. 
Prior work has also studied temporal models for $H_s$ and/or $T$ for fixed points in space. These models are often based on transformations of the marginal data to Gaussianity such that the temporal correlation can be modeled by ARMA-processes \citep[and the reference within]{lit:monbet2}. 
As stated above, we are instead interested in spatial models for $H_s$ and $T$, which for example are important when considering moving ships where the wave state at points visited on the ships route will be highly dependent. An important property of a spatial model for any larger region is that it allows for spatial non-stationarity \citep{lit:baxevani3, lit:ailliot}, i.e., different distributional behavior depending on the spatial location. 
Some prior work on modeling $H_s$ spatially, or spatio-temporally, using transformed Gaussian random fields exist. Such spatial models are usually based on a chosen parametric stationary covariance function for which parameters are estimated using maximum likelihood and/or minimum contrast methods. 
\citet{lit:baxevani2} considered regions small enough to assume stationarity in order to work with a stationary Gaussian model. To handle non-stationarity, this model was later extended in~\citet{lit:baxevani3} to a spatial moving average process with a non-stationary Gaussian kernel. 
\citet{lit:ailliot} instead considered mutually exclusive subregions of the spatial domain for which they assumed stationarity within. The mean and variance were estimated for each subregion and the measured values were standardized based on these parameters. The standardized data were then treated as stationary.

In \citet{lit:hildeman} a non-stationary and anisotropic model was proposed based on the SPDE approach \citep{lit:lindgren} and the deformation method \citep{lit:sampson}. 
Compared to the covariance-based models of \citep{lit:baxevani2, lit:baxevani3, lit:ailliot} this model is based on a description of the random field through a stochastic partial differential equation (SPDE).
By approaching the characterization of the random field from a SPDE perspective the model gains some distinct benefits. 
It allows modeling on complex spatial domains (even arbitrary Rimennian manifolds), a finite-dimensional representation of a continuously indexed Gaussian random field, and it has computationally beneficial properties (especially when modeling large regions).

The model we propose is an extension of the $H_s$ model by \citet{lit:hildeman}. Specifically, we will assume that the distribution of $H_s$ and $T$ are Gaussian after logarithmic transformation, as proposed by \citet{lit:ochi}. We will then model $\log(H_s)$ and $\log(T)$ using a bivariate extension of the model by \citet{lit:hildeman} where we also allow for  arbitrary smoothness of the two random fields as well as a spatially varying cross-correlation of the two quantities. 


%
%
The proposed model is not temporal and hence it cannot model the vast variability in sea state behavior over the whole year. Instead we restrict ourselves to modeling of the sea state variability during only one of the months of the year. The idea being that during a fixed month, the spatial sea state distribution does not change.
39 years of data from the north Atlantic during April month will be used to estimate the model as well as to validate it. 
To illustrate the flexibility of the proposed model, we will consider two safety issues in naval logistics which require spatial modeling of the sea state parameters, namely fatigue damage modeling of ships as well as estimation of the risk of capsizing due to broaching-to.




The structure of the paper is as follows.
In Section~\ref{sec:model}, the proposed model is introduced. Section~\ref{sec:modelDiscretization} describe the finite-dimensional discretization of the proposed model.
In Section~\ref{sec:data}, the data used for parameter estimation and validation of the model is described.
Section~\ref{sec:estimation} goes through the method of estimating the parameters of the model from the available data. 
It also assesses the fit of the model. 
Section~\ref{sec:applications} introduces two applications where such a spatial model can be used to estimate risks and wear associated with a planned ship journey. 
Finally, Section~\ref{sec:discussion} concludes with a discussion of the results and future extensions.

\section{Model formulation}
\label{sec:model}

In \citet{lit:hildeman} a random field model was developed for the significant wave height, $H_s$. The model was defined by interpreting $X(\psp) = \log(H_s)$ as a weak solution to the \textit{stochastic partial differential equation} (SPDE)
\begin{align}
\mathcal{L}^{\alpha/2} \left( \tau(\psp)\rv(\psp) \right) := \left[ \kappa(\psp)^{\frac{2}{\alpha}-2} \left( \kappa(\psp)^2 - \nabla \cdot  H(\psp) \nabla \right)  \right]^{\alpha/2}  \left(\tau(\psp)\rv(\psp)\right)  &= \noise(\psp).
\label{eq:SPDEG}
\end{align}
Here, $H$ is a symmetric and positive definite matrix-valued function, $\damp$ and $\tau$ are strictly positive real-valued functions, and $\alpha\ge 1$ a constant.
The SPDE is defined over a spatial domain, $\gspace$, and $\noise$ is Gaussian white noise.

When $\gspace := \mathbb{R}^d$, $\kappa(\psp) := \kappa>0$, $\tau(\psp) := \tau>0$, and $H(\psp) := I$ (the identity matrix),  the solution to \eqref{eq:SPDEG} is a mean-zero Gaussian random field with a Mat\'ern covariance function \citep{lit:whittle}. The parameters $\tau$ and $\kappa$ respectively controls the variance and correlation range of the field, and $\alpha = \nu + d/2$ where $\nu$ determines the smoothness. However, to obtain a model that is flexible enough to describe a wide range of non-stationary and anisotropic Gaussian random fields, the parameters $\kappa(\psp)$ and $H(\psp)$ of the model were obtained using the deformation method of \citet{lit:sampson} and the SPDE description of a Gaussian random fields with Mat\'ern correlation structure \citep{lit:whittle, lit:lindgren}.
In short that means that we consider a differentiable and bijective mapping, $\warp^{-1}(\psp)$, that maps points on the observational domain, $\gspace$, to points on a subset to some manifold, $\dspace$. When $X(\psp)$ is mapped to $\dspace$ it will be distributed as a Gaussian Mat\'ern field. Specifically, $\tilde{X}(\tilde{\psp}) := X(\warp^{-1}(\psp))$ is a unit-variance Gaussian random field with a Mat\'ern covariance function with the same smoothness parameter $\alpha$ as in \eqref{eq:SPDEG}. Because of this, the function $\warp$ explains the anistropy, non-stationarity, and correlation range of $X$, whereas $\tau(\psp)$ determines the marginal variances and $\alpha$ the smoothness. 

%
%
%
%
%
%
The connection between the parameters $H(\psp)$ and $\damp(\psp)$ of the SPDE in Equation \eqref{eq:SPDEG} and the mapping $\warp: \dspace \mapsto \gspace$ is
\begin{align}
    \damp^2(\psp) = \determinant{J[\warp^{-1}](\psp)}, \quad H(\psp) = \damp^2(\psp) J[\warp^{-1}]^{-1}(\psp) J[\warp^{-1}]^{-T}(\psp),
\end{align}
where $J[F^{-1}]$ denotes the Jacobian matrix of $F^{-1}$.
This means that the SPDE is completely characterized by the Jacobian matrix of $F$. 
In fact, the model is well-defined for a broader class than those which are diffeomorphic to a Mat\'ern Gaussian random field---it is enough that they are locally diffeomorphic to a Mat\'ern Gaussian random field. That is, any $d\times d$ matrix-valued function which is Lipschitz continuous and uniformly positive definite (or uniformly negative definite) can be used in place of $J[F^{-1}]$. 

\begin{figure}[t]
	\centering
	\begin{subfigure}{0.49\textwidth}
		\includegraphics[width=\textwidth]{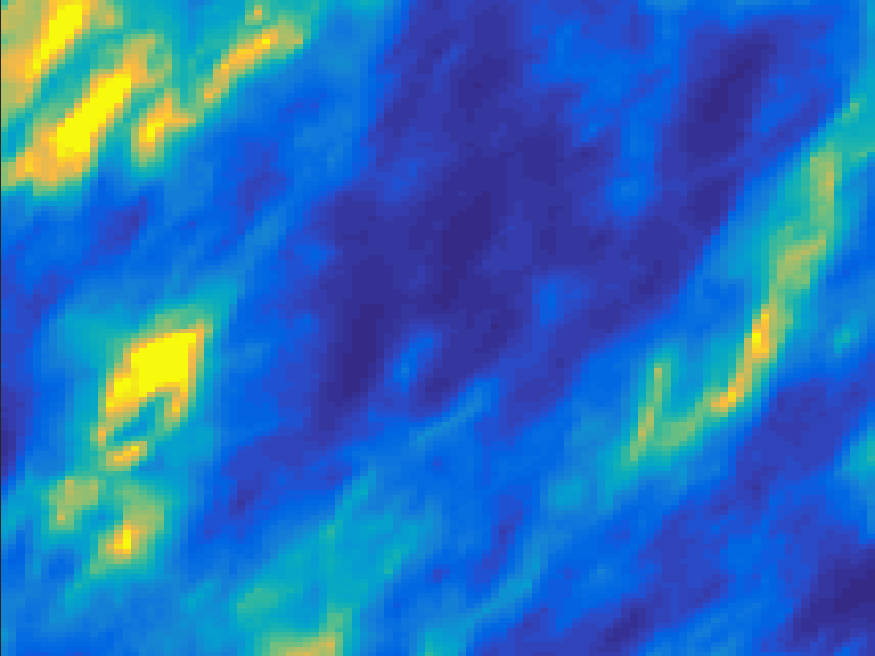}
	\end{subfigure}
	\begin{subfigure}{0.49\textwidth}
		\includegraphics[width=\textwidth]{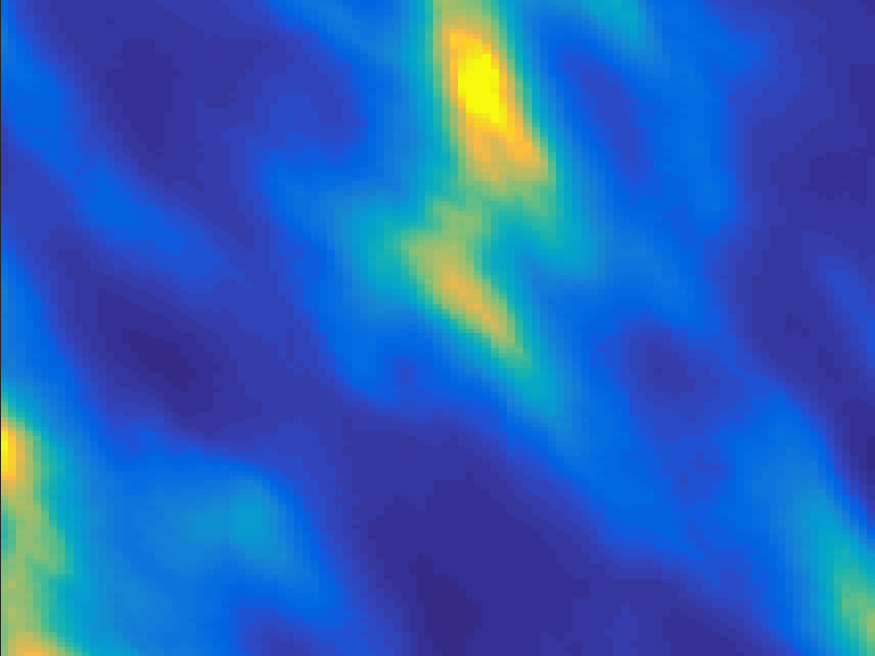}
	\end{subfigure}
	\caption{Realization of a bivariate, anisotropic and stationary Gaussian random field. The left field has a correlation range of $25$ in the direction of the principal axis at $45^\circ$ and a correlation range of $14$ in the perpendicular direction. The right field has the principal direction at an angle of $-45^\circ$ with the correlation range $30$, the perpendicular direction has a range of $15$. The correlation between the fields is controlled by $\rho = -0.98$. Furthermore, the left field has a smoothnes constant of $\alpha = 1.6$ while the right field has $\alpha = 3$.} 
	\label{fig:exAnisotropy}
\end{figure}

In \citet{lit:hildeman} it was shown that this SPDE model agreed well with data of significant wave height in the north Atlantic ocean.
We now extend the model to a bivariate random field model for significant wave height and wave period. 
We construct a bivariate model for which the marginal distributions over $H_s$ and $T$ are identical to the model of Equation \eqref{eq:SPDEG}. 
Let us denote $X(\psp) := \log H_s(\psp)$ and $Y(\psp) := \log T(\psp)$, and consider $X$ and $Y$ as dependent Gaussian random fields.
\citet{lit:bolin, lit:hu, lit:hu2} developed multivariate models of Gaussian random fields based on a triangular system of SPDEs. Inspired by those models, we extend \eqref{eq:SPDEG} to a bivariate model 
\begin{align}
\begin{bmatrix}
g_{11} & g_{12} \\
0 & g_{22} 
\end{bmatrix}
\begin{bmatrix}
\mathcal{L}_X^{\smooth/2} &0 \\ 
0 &\mathcal{L}_Y^{\beta/2}
\end{bmatrix}
\begin{bmatrix}
X \\ 
Y
\end{bmatrix}
&=:
D
\begin{bmatrix}
\mathcal{L}_X^{\smooth/2} &0 \\ 
0 &\mathcal{L}_Y^{\beta/2}
\end{bmatrix}
\begin{bmatrix}
X \\ 
Y
\end{bmatrix}
=
\begin{bmatrix}
 \mathcal{W} \\ 
 \mathcal{V}
\end{bmatrix}.
\label{eq:multidimHu}
\end{align}
Here $\mathcal{W}$ and $\mathcal{V}$ are independent copies of Gaussian white noise on $\gspace$ and $g_{11}, g_{12}$, and $g_{22}$ are scalar-valued functions in $L^{\infty}(\gspace)$, where $g_{11}$ and $g_{12}$ are bounded away from $0$ such that $D$ is invertible. 
The pseudo-differential operators $\mathcal{L}_X$ and $\mathcal{L}_Y$ are defined as in Equation \eqref{eq:SPDEG} and control the marginal distributions of $X$ and $Y$ independently.
The term $g_{12}$ will introduce dependencies between $X$ and $Y$.
The inverse, $R = D^{-1}$ can be used to rewrite the system of SPDEs as
\begin{align}
\begin{bmatrix}
\mathcal{L}_X^{\smooth/2} &0 \\ 
0 &\mathcal{L}_Y^{\beta/2}
\end{bmatrix}
\begin{bmatrix}
X \\ 
Y
\end{bmatrix}
&=
R
\begin{bmatrix}
 \mathcal{W} \\ 
 \mathcal{V}
\end{bmatrix}
:=
\begin{bmatrix}
 h_{11} &  h_{12} \\
 0 &  h_{22}
\end{bmatrix} 
\begin{bmatrix}
 \mathcal{W} \\ 
 \mathcal{V}
\end{bmatrix},
\label{eq:multidim}
\end{align}
which corresponds to a linear model of coregionalization \citep{lit:bolin}. The parameters $h_{11}, h_{12}$ and $h_{22}$ are here functions of the spatial location, fully defined by the parameters in the elements of $D$. In particular, $h_{12}$ solely defines the dependency between the two fields. Moreover, considering only one random field at a time, they will have the same distribution as in the univariate case if $h_{11}(\psp)^2+h_{12}(\psp)^2 = h_{22}(\psp)^2 = 1 \forall \psp \in \domsp$. 
In the case of $D$ being constant, \citet{lit:bolin} gives a parametrization of $R$ using only one parameter, $\rho$, due to the sum to one constraint. The parameter $\rho \in \R$ controls the correlation between the fields $X$ and $Y$ but is in general not equal to the correlation. 
Using $\rho$, the parameters of $D$ and $R$ are fully identified as 
\begin{align}
R &= 
\begin{aligned}
\begin{bmatrix}
h_{11} & h_{12} \\
0&h_{22}
\end{bmatrix}
=
\frac{1}{\sqrt{1+\rho^2}}\begin{bmatrix}
1 & \rho \\
 0 & \sqrt{1+\rho^2}
\end{bmatrix}
\end{aligned}, 
\quad
&D = R^{-1} = 
\begin{aligned}
\begin{bmatrix}
g_{11} &
g_{12} \\
& g_{22} 
\end{bmatrix}
=
\begin{bmatrix}
\sqrt{1+\rho^2} & - \rho \\
 0 & 1
\end{bmatrix}
\end{aligned}.
\end{align}
We use this parameterisation, but extend the model by allowing $\rho$ to be a spatially varying function. Hence, the model we consider is
\begin{equation}\label{eq:final_model}
\begin{split}
    \sqrt{1+\rho^2} \mathcal{L}_{X}^{\alpha/2} X - \rho \mathcal{L}_{Y}^{\beta/2} Y &= \mathcal{W} \\
    \mathcal{L}_{Y}^{\beta/2} Y &= \mathcal{V}.
    \end{split}
\end{equation}
With this parameterization, the covariance operators for $X$ and $Y$ are $\mathcal{L}_X^{-\alpha}$ and $\mathcal{L}_Y^{-\beta}$ respectively, and the cross-covariance is  
$\rho(1+\rho^2)^{-1/2}\mathcal{L}_X^{-\alpha/2}\mathcal{L}_Y^{-\beta/2}.
$
In the case when the covariance operators for $X$ and $Y$ are the same and $\rho$ is constant, the correlation coefficient between the two fields is equal to $\frac{\rho}{\sqrt{1+\rho^2}}$ in the sense that it corresponds to the Pearson correlation coefficient between the two fields at any fixed point in $\gspace$.
In the general case, the interpretation of $\rho$ as controlling the correlation still holds and 
values near zero of $\rho$ give a negligible dependency between the fields while large positive values give a strong positive correlation and large negative values give a strong negative correlation. However, a simple relationship with the pointwise correlation coefficient does not exist.
This effect is highlighted in Figure~\ref{fig:exAnisotropy} showing a realization of such a bivariate Gaussian random field model. Here, both fields are stationary and anisotropic but with different directions of the main principal axes and different smoothness parameters.
Even though $\rho = -0.98$, which would correspond to a correlation of $-0.7$ if the marginal random fields would have been equal in distribution, the true correlation between the fields is larger. It is however visible that peaks in the left field tend to correspond to valleys in the right field indicating a negative correlation.

\section{Model discretization}
\label{sec:modelDiscretization}

To be able to use the model of the previous section in applications, we first must discretize it. This is done using a finite element  approximation of the system of SPDEs. In this section we provide the details of this procedure. We first show the details in the univariate case with $\alpha=2$, then generalize to arbitrary $\alpha>1$, and finally combine the methods for the multivariate setting. 

\subsection{The univariate case}
\label{sec:weak}
In the case when $\alpha=2$ in \eqref{eq:SPDEG}, the model can be discretized using a standard Galerkin finite element method as suggested by \citet{lit:lindgren}. The aim is to approximate the solution $X$ by a basis expansion $X_h(\mv{s} ) = \sum_{j=1}^N U_j \phi_j(\mv{s})$. Here $\{\phi_j \}_{j=1}^N$ is a set of piecewise linear functions induced by a triangular mesh of the spatial domain. Let $V_h$ be the space spanned by these basis functions. Augmenting the operator with homogeneous Dirichlet boundary conditions and considering the weak formulation of the SPDE on $V_h$ yields the following system of equations for the coefficients in the basis expansion
\begin{align}
\sum_{j=1}^N \left( \langle \damp \tau \phi_j , \phi_i \rangle 
+ \langle  H \nabla \tau \phi_j, \nabla \left(\damp^{-1} \phi_i\right) \rangle  \right) U_j \overset{d}{=} \langle \noise, \phi_i \rangle, \quad i=1,\ldots, N,
\end{align}
where $\langle \cdot, \cdot \rangle$ denotes the inner product on $\mathcal{G}$. This system of equations can be written in matrix form as $KU := (B + G)U \overset{d}{=} W$, where 
$B_{ij} := \langle \kappa  \phi_j, \phi_i \rangle$, $G_{ij} := \langle  H \nabla  \phi_j, \nabla \left( \damp^{-1} \phi_i \right) \rangle$, and $W \sim \mathbb{N}(0, C)$ with $C_{ij} := \langle \phi_j, \phi_i \rangle$. Hence, the stochastic weights of the basis expansion are $U \sim \mathbb{N}(0, K^{-1}CK^{-T})$.

The important property of using a basis of $V_h$ with compact support is that $K$ and $C$ will be sparse matrices. \citet{lit:lindgren} showed that $C$ can be approximated by a diagonal matrix, with diagonal elements $\langle \phi_i, 1 \rangle$. With this approximation, the precision matrix $KC^{-1}K$ is also sparse and $U$ is Gaussian Markov random field (GMRF). This greatly reduces the computational cost for inference and simulation \citep{lit:rue}. We refer to \citep{lit:hildeman} for further details in the univariate case. 

\subsection{Rational approximation for arbitrary smoothness}
\label{sec:rational}
The procedure from the previous subsection can be extended to integer values of $\alpha$ by noting that the solution to $\mathcal{L}^2 X = \noise$ can be obtained by first solving $\mathcal{L} X_1 = \noise$ and then $\mathcal{L} X = X_1$. One can therefore use the discretization from the previous subsection iteratively to obtain a discretization for even integer values of $\alpha$. \citet{lit:lindgren} also stated the solution to $\mathcal{L}^{1/2}X = \noise$ as a least square solution, which can be combined with the iterative procedure to obtain discretizations also for odd integer values of $\alpha$. This was utilized in \citep{lit:hildeman} where only integer values of $\smooth$ were considered. 

For large values of $\alpha$, the correlation function does not change much for a small change in $\alpha$. However, for small values of $\alpha$, restricting it to integer values constrain the flexibility of the model. For instance, the exponential correlation function corresponds to $\alpha = 1.5$ and cannot be modeled by an integer-valued $\alpha$. 
Therefore, in this work we want to model any positive value of $\alpha \ge 1$ and not only integer values. 
Until recently, it was not clear how to formulate a FEM approximation for non-integer valued $\alpha$. However, \citet{lit:bolin3} solved this problem by combining the FEM approximation with a rational approximation of the power function, i.e., $x^{\smooth} = \frac{p_l(x)}{p_r(x)}$, where $p_l$ and $p_r$ are polynomials of some chosen orders. 
By using such a decomposition, it was possible to approximate the non-integer power of a pseudo-differential operator $\mathcal{L}^{\smooth}$ as a product of two polynomial pseudo-differential operators, $P_l$ and $P_r$. Here, $P_l = \sum_{j=0}^{N_r} a_j \mathcal{L}^{j}$ and similarly for $P_r$.
That is, 
\begin{align}
\mathcal{L} \rv^R_m \approx \left( P_r^{-1}P_l \right)\rv^R_m = \noise \Leftrightarrow P_{l} \rv^{R}_ m = P_r \noise,
\label{eq:rationalApprox}
\end{align}
where $\rv^R_m$ is the rational approximation of $\rv$. 
Since the polynomial operators $P_l$ and $P_r$ are commutative, the solution can be written as a system of equations
\begin{align}
    &P_l Z = \noise \\
    &X^R_m = P_r Z.
\end{align}
This is important since a FEM approximation of $P_l Z = \noise$ can be used in order to get a GMRF approximation of $Z$. 
More specifically, the discretized FEM operators $P_l$ and $P_r$ can be written as 
\begin{align}
P_{l,h} := b_{m+1}K^{m_{\alpha}-1} \prod_{j=1}^{m+1} \left( I - r_{2,j} K \right), &\quad 
P_{r,h} := c_m \prod_{i=1}^{m} \left( I - r_{1,i} K \right),
\end{align}
where $K$ is the FEM matrix of Section \ref{sec:weak}, $m$ is an integer controlling the quality of the approximation, and $m_{\smooth} = \max\{1, \lfloor \smooth \rfloor \}$ an integer associated with the smoothness parameter, $\smooth$. The coefficients $b_{m+1}, c_m, r_{2,j},r_{i,j}$ are obtained from the rational approximation of the function $x^{\smooth}$ (see \citet{lit:bolin3}). A larger $m$ yields a better approximation, $X^R_m$, but also more terms in the polynomial operators which will increase the computational cost by making $P_{l,h}$ and $P_{r,h}$ less sparse.

The distribution of the stochastic weights is $U\sim \mathbb{N}(0, P_{r,h} P_{l,h}^{-1} C P_{l,h}^{-T} P_{r, h}^T)$. Even though both $P_{r,h}$ and $P_{l,h}$ are sparse, their inverses are not. Therefore, the precision matrix of $U$ will not be sparse. However, because of the two-step procedure of the model formulation, all computational benefits of the GMRF case can be maintained when using the model. The trick is to use the nested SPDE approach \citep{lit:bolin4} and write $U = P_{r,h}\tilde{U}$, since $P_{r,h}$ is sparse and $\tilde{U}$ has a sparse precision matrix $P_{l,h}^{T} C^{-1} P_{l,h}$.

\subsection{FEM for the bivariate model}
We are now ready to discretize the model of Equation \eqref{eq:final_model}. In the prior section we saw that we can write a FEM approximation of the operator $\sqrt{1+\rho^2} \mathcal{L}_{X}^{\alpha/2}$ as $K_X = P_lP_r^{-1}$. Likewise, denote the FEM approximation of the operator $\mathcal{L}_{Y}^{\beta/2}$ as $K_Y = Q_l Q_r^{-1}$. Moreover, we can consider $-\rho \mathcal{L}_{Y}^{\beta/2}$ to be a composition of the two operators $-\rho \mathcal{I}$ and $\mathcal{L}_{Y}^{\beta/2}$. By considering an iterative FEM approximation with respect to these two operators we acquire the system of linear equations
\begin{align}
     K_X U_X + K_{\rho} U_Y &= W \\
    K_Y U_{Y} &= V,
\end{align}
where $W$ and $V$ are i.i.d.~$\mathbb{N}(0, C)$ random vectors and $U_X$ and $U_Y$ are the stochastic weights for the FEM approximation of $X$ and $Y$ respectively. Furthermore, $K_{\rho} = -C^{-1}C_{\rho}K_Y$ where $C_{\rho} = \{\langle \rho(\psp) \phi_i(\psp), \phi_j(\psp) \rangle\}_{ij}$.
The block covariance matrix for $U_X$ and $U_Y$ is
\begin{align}
    &\begin{bmatrix}
    \sigma_X & \sigma_{XY} \\
    \sigma_{YX} & \sigma_{Y}
    \end{bmatrix} =
    \begin{bmatrix}
    K_X^{-1} C K_X^{-T} + K_X^{-1} K_{\rho} K_Y^{-1} C K_Y^{-T} K_{\rho}^{T}K_X^{-T}
    & -K_X^{-1}K_{\rho} K_Y^{-1} C K_Y^{-T} \\
    - K_Y^{-1} C K_Y^{-T} K_{\rho}^T K_X^{-T}   
    & K_Y^{-1} C K_Y^{-T}
    \end{bmatrix}.
    \end{align}
    
    The corresponding block precision matrix is
    \begin{align}
    &\begin{bmatrix}
    q_X & q_{XY} \\
    q_{YX} & q_{Y}
    \end{bmatrix} =
    \begin{bmatrix}
    K_X^T C^{-1} K_X & 
    K_X^T C^{-1}  K_{\rho} \\
    K_{\rho}^T C^{-1} K_X
    & K_Y^T C^{-1} K_Y +  K_{\rho}^T C^{-1} K_{\rho} 
    \end{bmatrix}.
    \end{align}

    Note that this is not a sparse matrix, which is needed to acquire the important computational advantages of the SPDE approach. However, by using the idea  introduced in the previous section, we can formulate the model as a latent GMRF to keep the computational benefits. This is done by 
    considering $U_X = P_r \tilde{U}_X$ and $U_Y = Q_r \tilde{U}_Y$ where $\left[ \tilde{U}_X, \tilde{U}_Y \right] \sim \mathbb{N}\left( \bs{0}, \tilde{Q}\right)$ is a GMRF with 
    \begin{align}
    \tilde{Q} &=
    \begin{bmatrix}
    P_l^T C^{-1} P_l & 
    -P_l^T C^{-2}C_{\rho}Q_l \\
    -Q_l^T C_{\rho}^T C^{-2} P_l  
    & Q_l^T \left( C^{-1} + C_{\rho}^T C^{-3} C_{\rho} \right) Q_l
    \end{bmatrix}.
    \end{align}
    With this formulation of our model, we can use the methods of \citet{lit:bolin3} for computationally efficient inference and simulation.


\section{Data}
\label{sec:data}
In order to test the proposed model, we will fit it to data from the ERA-Interim global atmospheric reanalysis \citep{lit:dee} acquired by the European Centre for Medium-Range Weather Forecasts (ECMWF).
The reanalysis data is based on measurements and interpolated to a lattice grid in a longitude-latitude projection using ECMWFs weather forecasting model IFS, cycle 31r2 \citep{lit:berrisford}.
The spatial resolution of the data is $0.75^{\circ}$ and it is available from 1979 to present. 
We will use the variables \textit{significant wave height of wind and ground swells} and \textit{mean wave period} from the dataset as $H_s$ and $T_1$ in our analysis. Both variables are available at a temporal resolution of 6 hours. However, since we will not model the temporal evolution of the data, and therefore want to approximate data from different points in time as independent, we thin the data to a temporal resolution of $24$ hours. Data from different months are distributed differently due to the effects of the annual cycle. Because of this, we restrict the analysis to the data from the month of April for the available years 1979 to 2018.

\begin{figure}[t]
	\centering
	\begin{subfigure}{0.49\textwidth}
	\centering		
	\includegraphics[width=\textwidth]{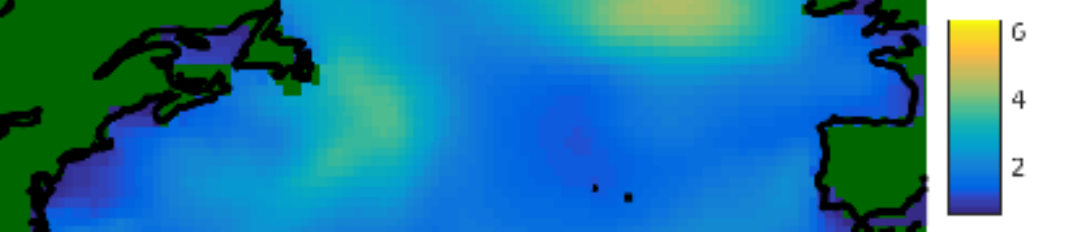}
	\caption{$H_s$ data $1$}		
	\end{subfigure}
	\begin{subfigure}{0.49\textwidth}
	\centering		
	\includegraphics[width=\textwidth]{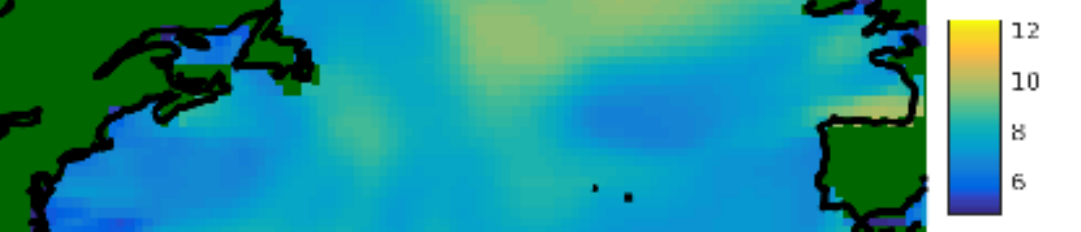}	
	\caption{$T_1$ data $1$}		
	\end{subfigure}		\\
    \begin{subfigure}{0.49\textwidth}
	\centering		
	\includegraphics[width=\textwidth]{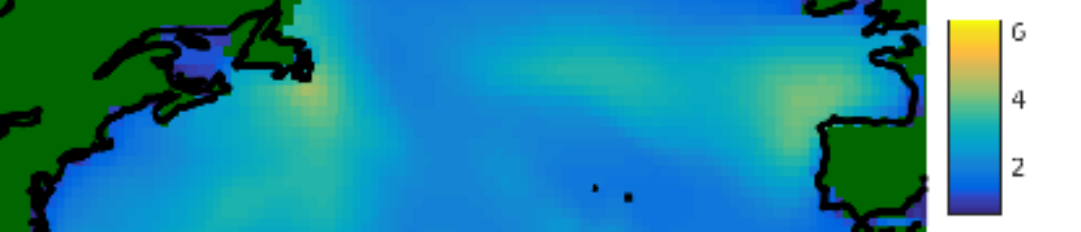}
	\caption{$H_s$ data $2$}		
	\end{subfigure}
	\begin{subfigure}{0.49\textwidth}
	\centering		
	\includegraphics[width=\textwidth]{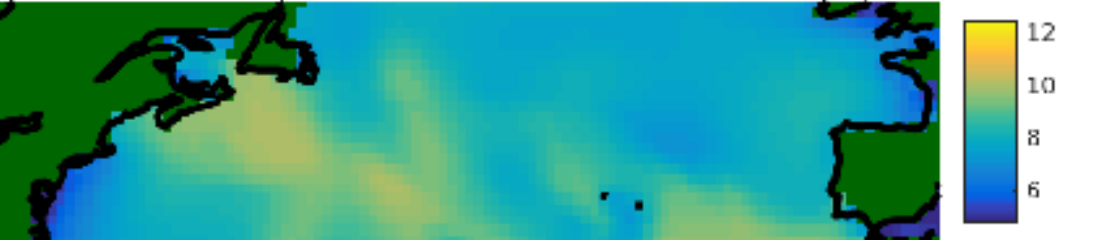}	
	\caption{$T_1$ data $2$}		
	\end{subfigure}		
	\caption{Two observations of $H_s$ and $T_1$ chosen randomly from the dataset of April month during the years 1979-2018.}
	\label{fig:realizations}
\end{figure}

\begin{figure}[t]
\centering
\begin{subfigure}{0.32\textwidth}
\includegraphics[width = \textwidth, keepaspectratio]{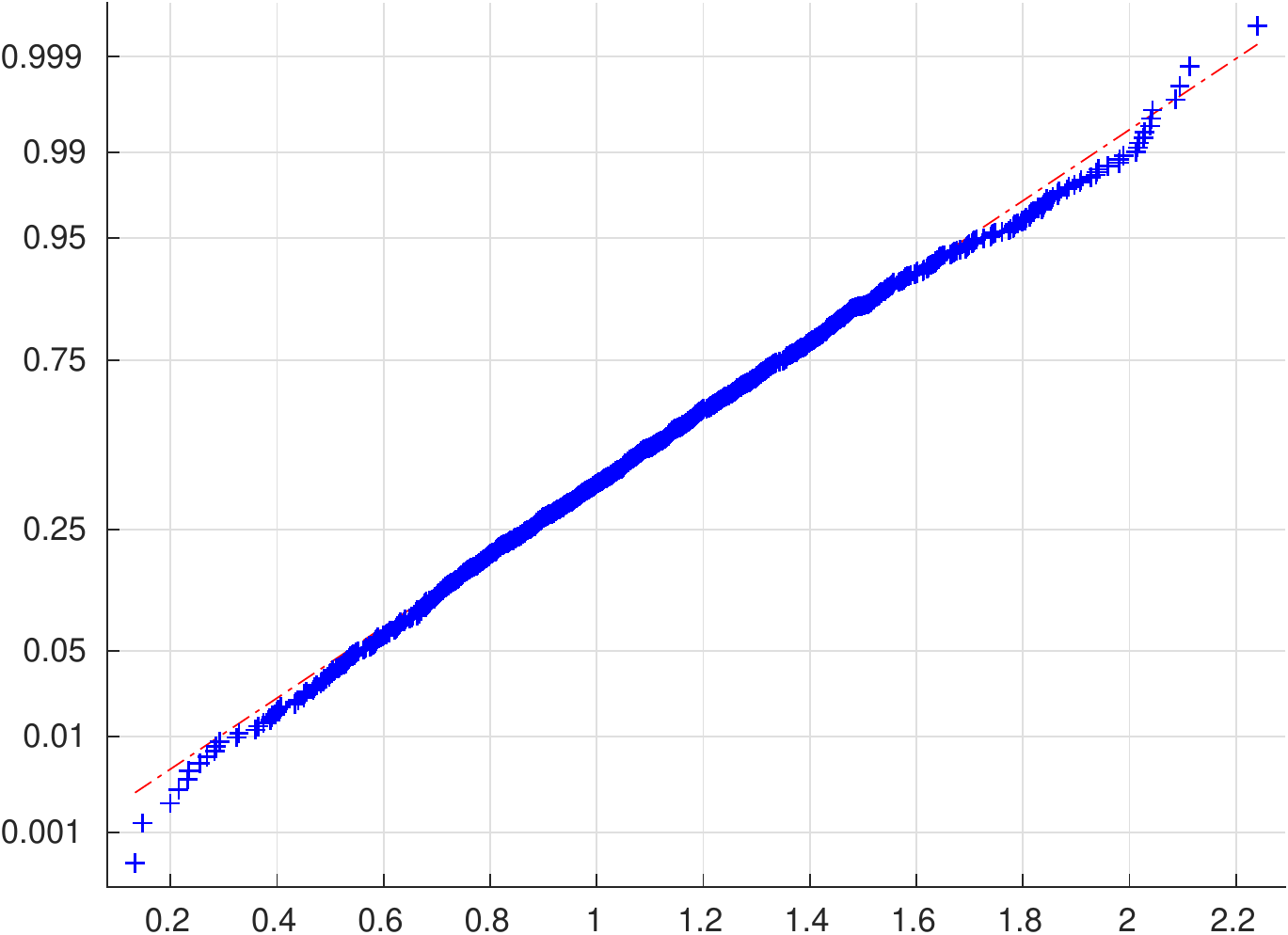}
\caption{$\log H_s$}
\end{subfigure}
\begin{subfigure}{0.32\textwidth}
\includegraphics[width = \textwidth, keepaspectratio]{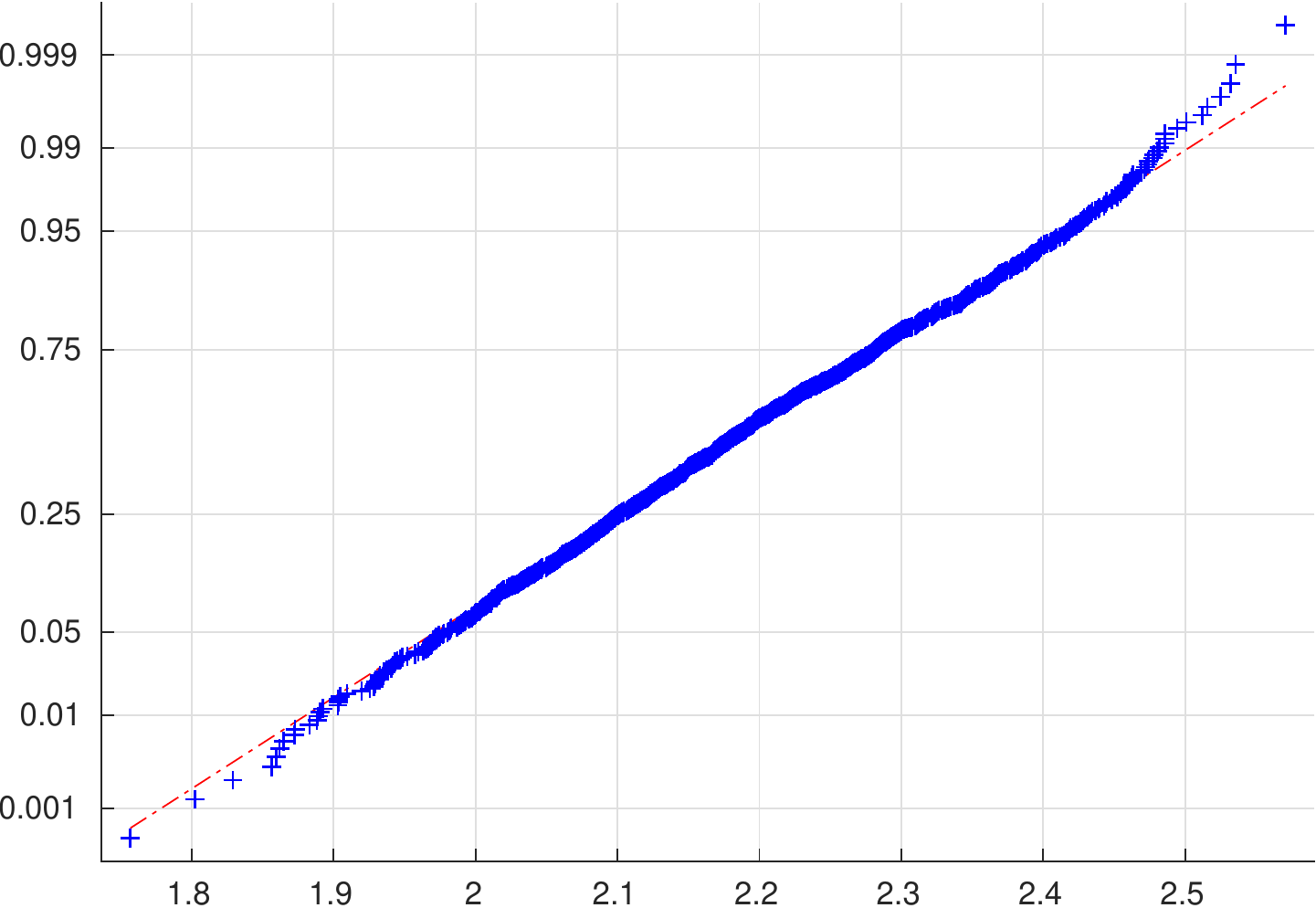}
\caption{$\log T_1$}
\end{subfigure}
\begin{subfigure}{0.32\textwidth}
\includegraphics[width = \textwidth, keepaspectratio]{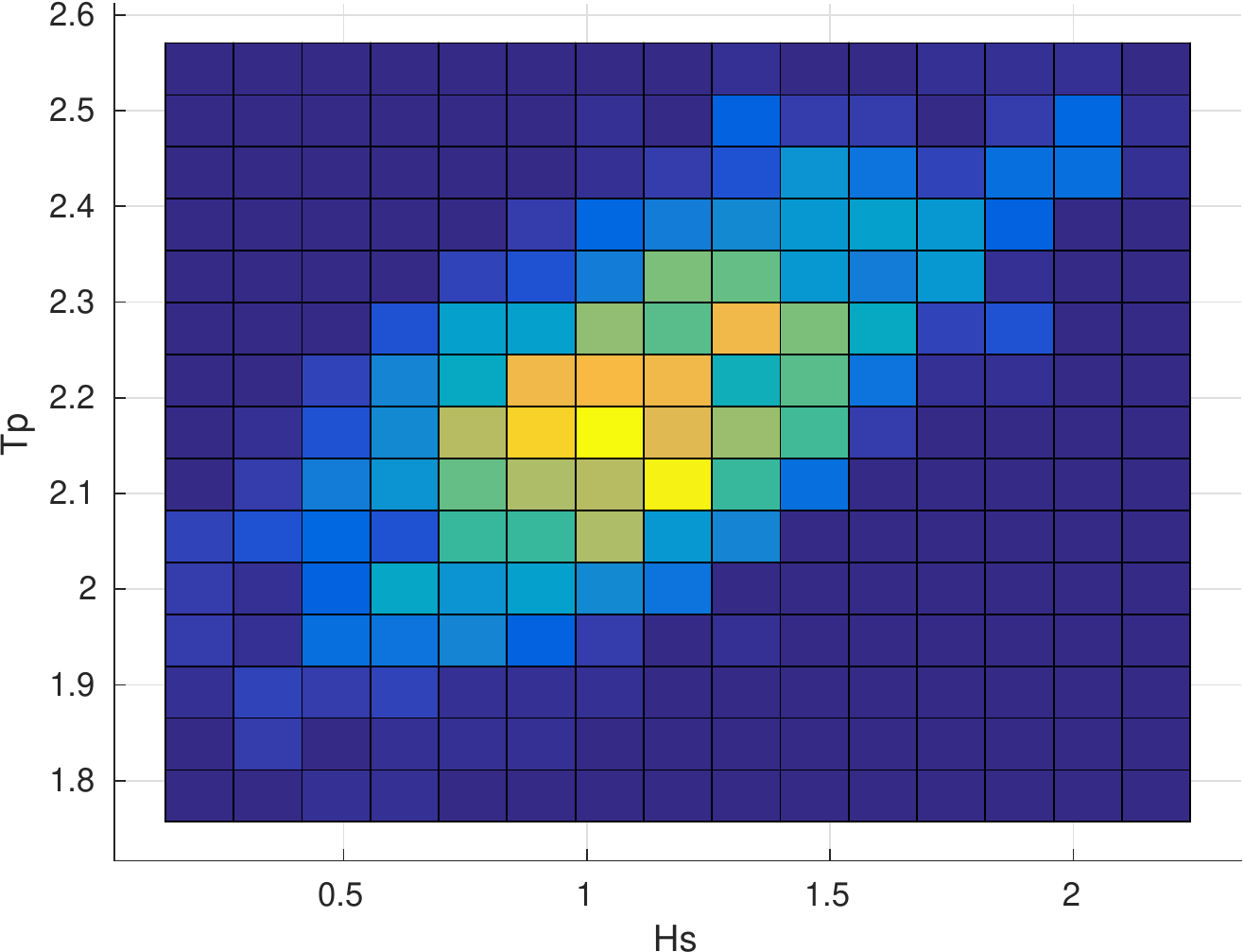}
\caption{Bivariate histogram}
\end{subfigure}
\caption{Normal probability plots of the marginal distribution of $\log H_s$ and $\log T_1$ as well as their corresponding two dimensional histogram. The data is taken from a point at latitude $48.75^\circ$ and longitude $-35.25^\circ$ from the ERA-Interim dataset.}
\label{fig:normplots}
\end{figure}

\begin{figure}[t]
	\centering
	\begin{subfigure}{0.4\textwidth}
	\centering		
	\includegraphics[width=\textwidth]{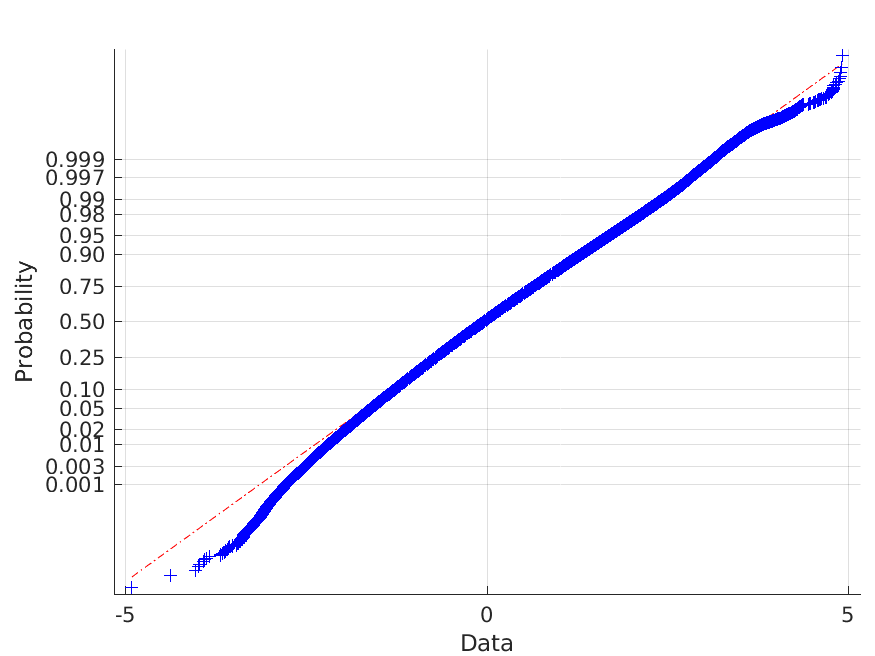}
	\caption{$\log H_s$}		
	\end{subfigure}
	\begin{subfigure}{0.4\textwidth}
	\centering		
	\includegraphics[width=\textwidth]{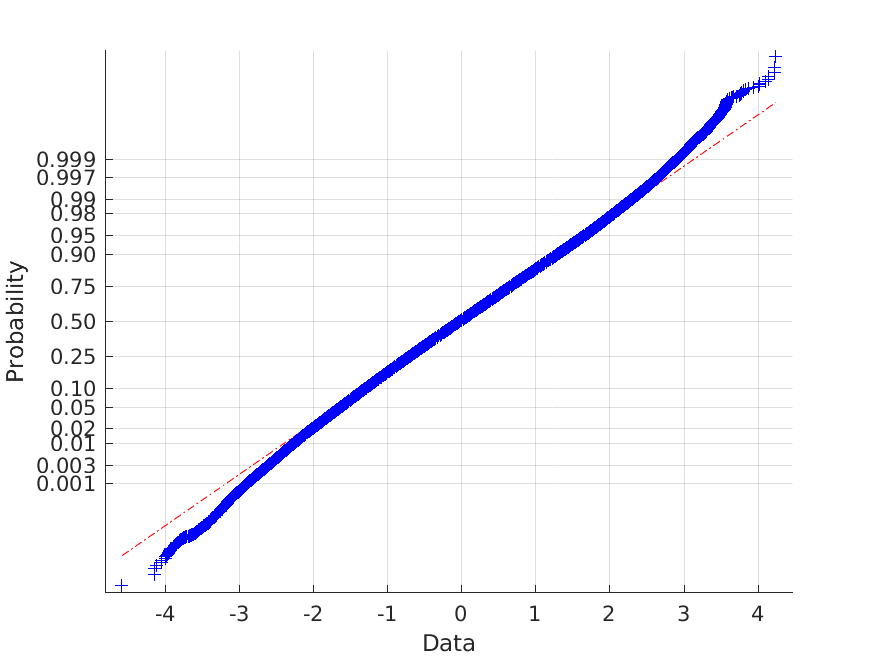}	
	\caption{$\log T_1$}		
	\end{subfigure}		
	\caption{Normal probability plot of all data (standardized for each spatial location separately prior to computing the normal probability plot). Left: plot for $\log H_s$. Right: Plot of $\log T_1$.}
	\label{fig:dataNormplots}
\end{figure}

We also restrict the analysis spatially to the north Atlantic, since this region contains several important trading routes and is known to produce data that is approximately log-Gaussian distributed \citep{lit:ochi}. 
An example of two simultaneous observation of $H_s$ and $T_1$ from the data can be seen in Figure \ref{fig:realizations}.
A bivariate histogram as well as marginal normal distribution plots for $\log H_s$ and $\log T_1$ for one specific point in space ($-32.25^{\circ}$ longitude and $48.75^{\circ}$ latitude) can be seen in Figure \ref{fig:normplots}. The data at this point agrees well with the assumption of a bivariate log-normal distribution, and similar results are obtained for  other locations in the domain. 
Figure \ref{fig:dataNormplots} shows the normal probability plot of $\log H_s$ and $\log T_1$ over all points in the region. The data were first standardized, pointwise, before computing the plot. Hence, the points should lie on a line if the assumption of log-normality holds, which can be seen to be true in the figure.

\begin{figure}[t]
	\centering
	\begin{subfigure}{0.49\textwidth}
		\centering
		\includegraphics[width = \textwidth]{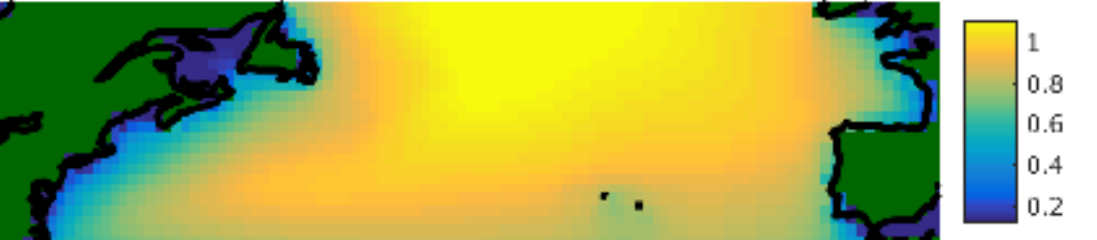}
		\caption{mean $H_s$}
	\end{subfigure}
	\begin{subfigure}{0.49\textwidth}
		\centering
		\includegraphics[width = \textwidth]{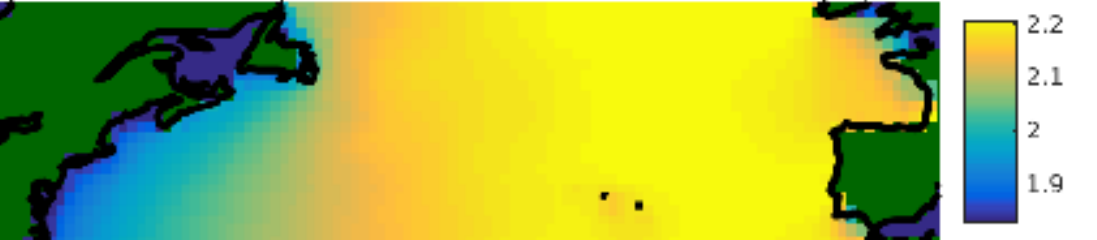}
		\caption{mean $T_1$}
	\end{subfigure}
	\begin{subfigure}{0.49\textwidth}
		\includegraphics[width = \textwidth]{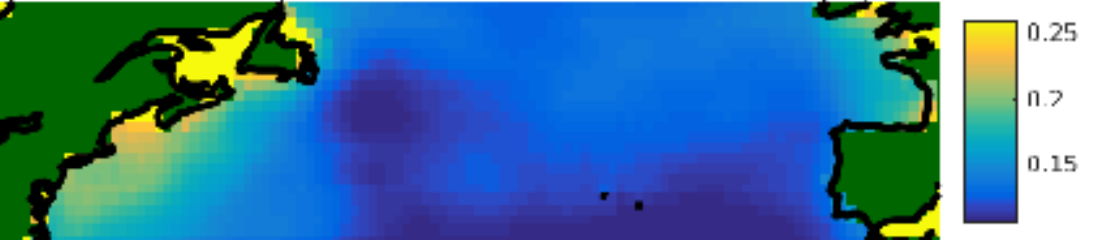}
		\caption{variance $H_s$}
	\end{subfigure}
	\begin{subfigure}{0.49\textwidth}
		\includegraphics[width = \textwidth]{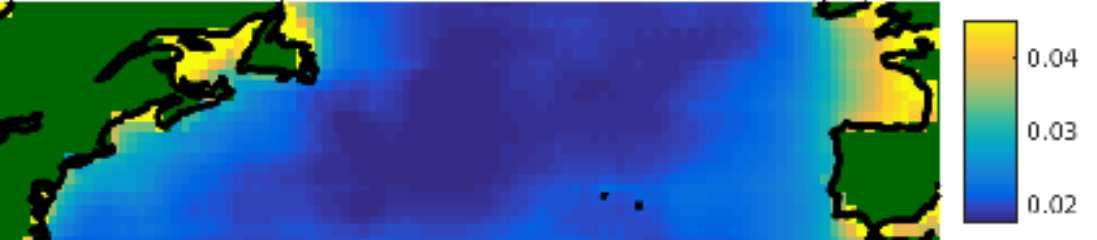}
		\caption{variance $T_1$}
	\end{subfigure}
	\caption{Sample mean and sample variance for both $H_s$ and $T_1$ in the north Atlantic. }
	\label{fig:meanVarFields}
\end{figure}

The sample mean and sample variance of the logaritmized data of April months can be seen in Figure \ref{fig:meanVarFields}. 
Clearly, the mean wave height is decreasing close to the coasts and the wave height variance is slightly increasing close to the coasts. The mean wave period is larger to the east than in the west. This is due to the mean wind direction blowing eastward. Also wave period show similar behavior.

The left columns of Figures \ref{fig:correlationHs} and \ref{fig:correlationTp} show the empirical correlation between three reference points in space and every other point in the spatial domain. 
Apparently, the point close to the coast of USA is showing an anisotropic pattern with the principal axis on the diagonal. Contrary to this, the spatial correlation of the mid Atlantic and at the coast of northern Europe has the principal axis in the east-west direction. 
It should be noted that the data is portrayed in the longitude-latitude coordinate system in Figures \ref{fig:correlationHs} and \ref{fig:correlationTp}. Other projections would yield different shapes of anisotropy---however, it is clear that no stationary model (on the sphere or in the plane) can explain the observed behaviour.

The considered dataset consists of 1200 days of data. We divide these into two equally-sized subsets of training data and test data. The training set consists of every second day starting from the first day available. The test set consists of the remaining days. Hence, the test- and training sets form a partition of all available days, each set consists of $600$ days, at least 2 days apart. In the next section we will use the training set to estimate model parameters. The test set is used to compare the fitted model with data for model validation. 

\section{Parameter estimation and model fit}
\label{sec:estimation}

\begin{figure}[t]
	\centering
	\includegraphics[width=0.5\textwidth]{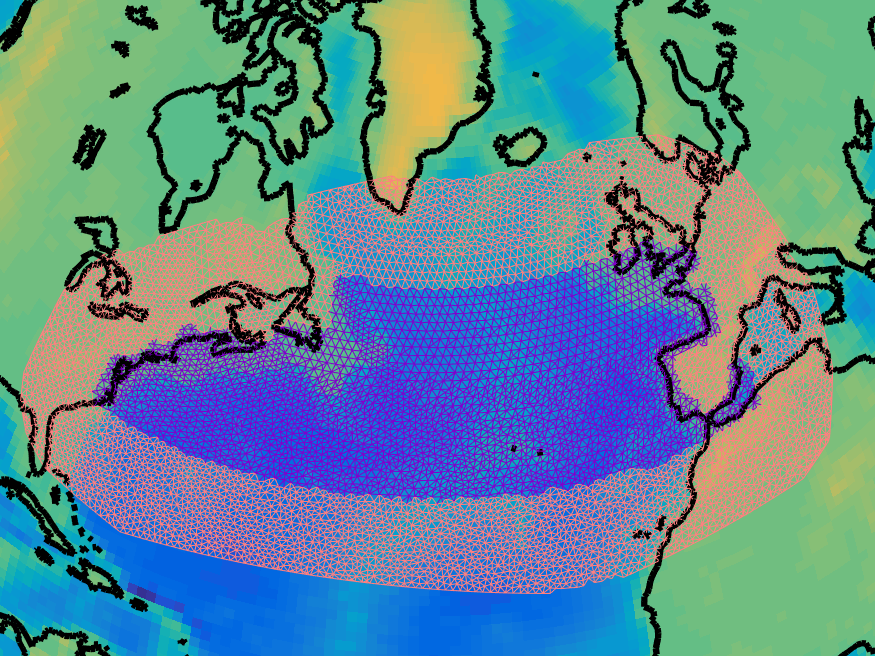}
	\caption{The north Atlantic with the the FEM mesh overlaid. Blue triangles are part of the spatial domain, $\gspace$. Pink triangles are part of the mesh extension. }
	\label{fig:globeMesh}
\end{figure}

Just as in \citet{lit:hildeman}, we logaritmize and standardize the data first, marginally pointwise using sample mean and sample variances from the training set. The standardized data is then modeled by the proposed mean-zero bivariate Gaussian random field where we fix the marginal variances to one. As is common in geostatistical models, we allow for a nugget effect for each dimension while estimating the model. That is, for a location $\mv{s}_i$, we assume that the observed values, $X_{obs,i}, Y_{obs,i}$, are $X_{obs,i} = X(\mv{s}_i) + \varepsilon_{X,i}$ and $Y_{obs,i} = Y(\mv{s}_i) + \varepsilon_{Y,i}$, where $\varepsilon_{X,i} \sim \mathbb{N}(0,\sigma_{X,e}^2)$ and $\varepsilon_{Y,i} \sim \mathbb{N}(0,\sigma_{Y,e}^2)$ are independent variables representing measurement noise.

In order to use the proposed FEM model, a triangular mesh has to be created over the spatial domain, $\gspace$. Since the spatial domain is in reality a subset of the surface of the globe---we create a mesh approximating $\gspace$ by a polyhedra, i.e., as a piecewise planar manifold. Hence, the region inside each triangle is planar. 
Figure \ref{fig:globeMesh} shows the mesh created for the north Atlantic. The blue triangles correspond to triangles within $\gspace$ and the pink triangles make up the mesh extension used to remove boundary effects. As in \citet{lit:hildeman}, the barrier method \citep{lit:bakka} is used to reduce the required size of the mesh extension.

Since the parameters of the proposed model are not known a priori, they have to be estimated from data. 
The proposed bivariate model is defined by the marginal random fields through $K_X$ and $K_Y$, and the cross-correlation function $\rho(\psp)$. 
The likelihood function of the joint model can be computed explicitly with a computational cost of $\mathcal{O}(N^{3/2})$, where $N$ are the number of nodes in the triangular mesh. 
The maximum likelihood (ML) estimates of the parameters cannot be computed explicitly, but instead numerical optimization using a quasi-Newton algorithm is used to acquire the parameter estimates. Furthermore, the initial values of the optimization algorithm is chosen using local parameter estimates as proposed in \citep{lit:hildeman}.

Although the joint likelihood can be optimized numerically, we propose a stepwise parameter estimation procedure, motivated as follows: 
One of the strengths of the proposed model is that all parameters have intuitive interpretations.
The parameters of $K_X$ and $K_Y$ respectively explain the spatial distribution of the random fields $X$ and $Y$ independently of each other.
Since the real spatial cross-correlation structure between $X$ and $Y$ likely is too complex to be explained completely by just $\rho(\psp)$, some degree of model-misspecification will be present. 
Maximizing the full likelihood function corresponds, asymptotically, to minimizing the Kullback-Liebler divergence between the true data distribution and the assumed model. However, under model-misspecification, full ML estimates of the bivariate fields do not necessarily estimate the parameters of the original interpretation; instead, the estimates will correspond to the values that are minimizing the distance between the true model and the proposed one. 
In many applications there is a point in keeping the original interpretation rather than minimizing the distributional distance---especially if conclusions should be drawn based on the estimated values of the parameters themselves.
Therefore, we fit $X$ and $Y$ independently in a first step. Then, conditioned on the estimates of the univariate random field parameters, a ML estimate of the cross-correlation structure, $\rho(\psp)$, is computed.

Estimating the parameters of $K_X$ and $K_Y$ independently has the additional advantage that it allows a lower dimensionality in the quasi-Newton optimization; which reduces the computational cost of estimation as well as decreases the risk of finding bad local optima. Also, the parameters of $K_X$ and $K_Y$ independently can be computed in parallel, further reducing the wall clock time.

\subsection{Estimation of the univariate random fields}

The models for $X$ and $Y$ independently are parametrized by the smoothness $\alpha$, the nugget effect, as well as the functions $H(\psp)$ and $\kappa(\psp)$. As in \citet{lit:hildeman}, we define 
\begin{align}
\tilde{H}(\psp) = \begin{bmatrix}
\exp\left(h_1(\psp)\right) & \left(2S(h_3(\psp)) - 1\right)\exp \left( \frac{h_1(\psp) + h_2(\psp)}{2} \right) \\ 
\left(2S(h_3(\psp)) - 1\right)\exp \left( \frac{h_1(\psp) + h_2(\psp)}{2} \right)  & \exp \left(h_2(\psp)\right)
\end{bmatrix},
\end{align}
and let $\kappa(\psp) = \determinant{\tilde{H}(\psp)}^{-1/2}$, and $H(\psp) = \kappa(\psp)^2 \tilde{H}$. The functions $h_1,h_2,h_3$ are defined as low-dimensional regressions on cosine functions over the domain of interest,
\begin{align}
h_i(\psp) = \sum_{p=0}^k \sum_{n=0}^k \beta_{np}^i \cos \left( n\frac{\pi s_1}{S_1} \right) \cos \left( p\frac{\pi s_2}{S_2} \right),\quad i=1,2,3,    
\label{eq:cosineParams}
\end{align}
where $\psp = (s_1,s_2)$ and $S_1, S_2$ denotes the width and height of the bounding box of the locations of observations. The advantage of this parameterization is that we do not have any restrictions on the coefficients $\beta_{np}^i$ in order to obtain a valid model.
We use $k = 4$ in Equation \eqref{eq:cosineParams}, meaning that $25\cdot 3 + 2 = 77$ parameters were estimated simultaneously using the quasi-Newton method for each field.

\begin{figure}[t]
	\centering
	\begin{subfigure}{0.9\textwidth}
	\centering
    	\begin{subfigure}{0.49\textwidth}
    		\centering
    		Data $\log H_s$
    	\end{subfigure}
    	\begin{subfigure}{0.49\textwidth}
    		\centering
    		Model $\log H_s$
    	\end{subfigure}	
    	\begin{subfigure}{0.49\textwidth}
    		\centering
    		\includegraphics[width = \textwidth]{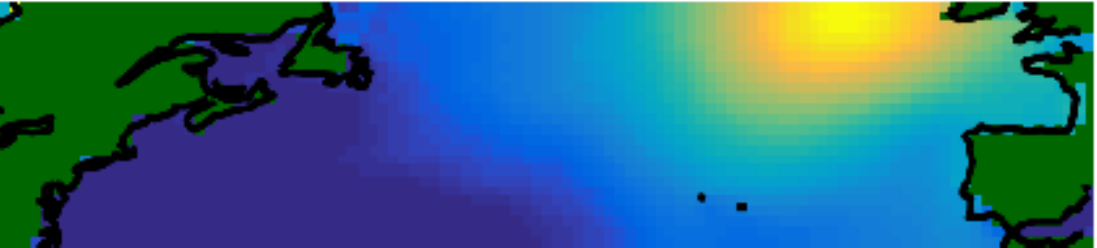}
    	\end{subfigure}
    	\begin{subfigure}{0.49\textwidth}
    		\includegraphics[width = \textwidth]{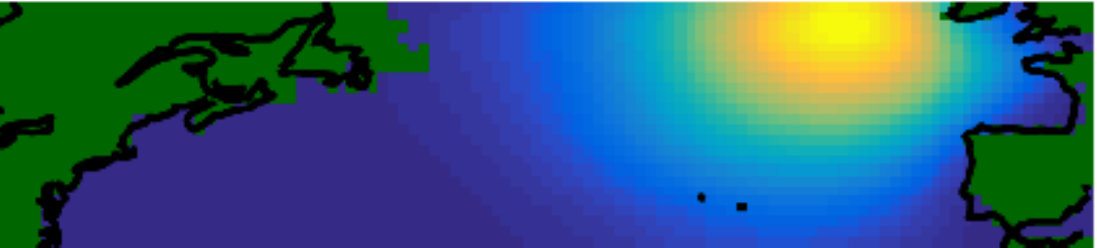}
    	\end{subfigure}
    	\begin{subfigure}{0.49\textwidth}
    		\centering
    		\includegraphics[width = \textwidth]{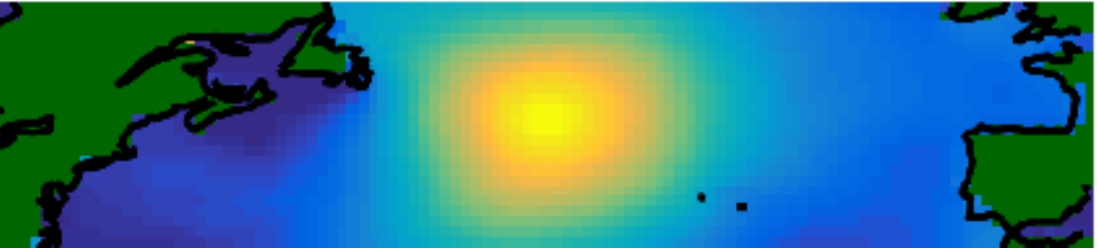}
    	\end{subfigure}
    	\begin{subfigure}{0.49\textwidth}
    		\includegraphics[width = \textwidth]{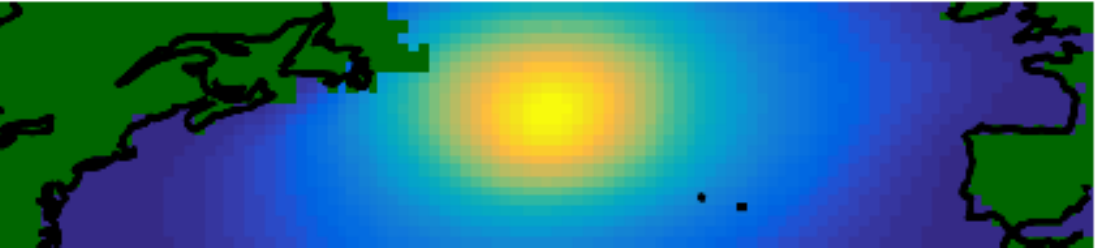}
    	\end{subfigure}
    	\begin{subfigure}{0.49\textwidth}
    		\centering
    		\includegraphics[width = \textwidth]{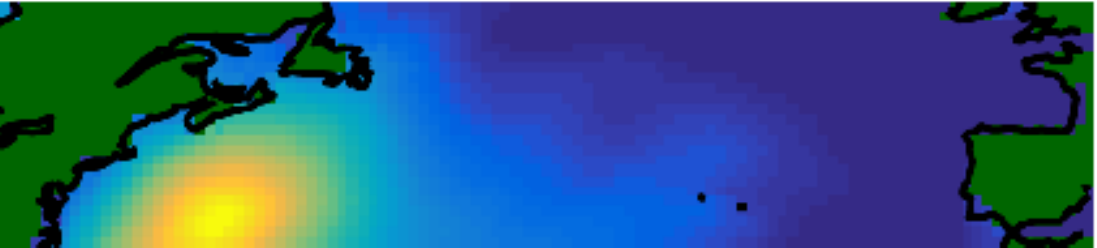}
    	\end{subfigure}
    	\begin{subfigure}{0.49\textwidth}
    		\includegraphics[width = \textwidth]{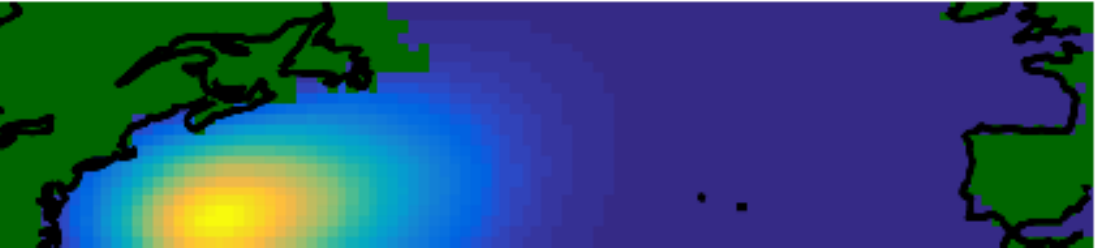}
    	\end{subfigure}	
	\end{subfigure}
	\begin{subfigure}{0.07\textwidth}
	\centering
	\includegraphics[width=\textwidth]{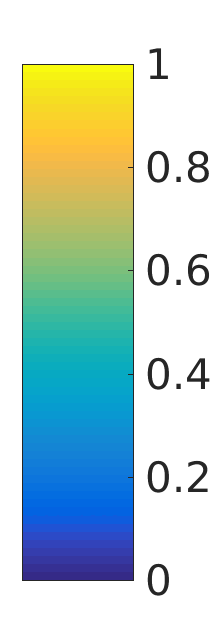}
	\end{subfigure}		
	\caption{Correlation between three different reference points and all other points in $\log H_s$. Left column: empirical correlation function from data. Right column: correlation function from fitted model. }
	\label{fig:correlationHs}
\end{figure}

\begin{figure}[t]
	\centering
	\begin{subfigure}{0.9\textwidth}
	\centering
    	\begin{subfigure}{0.49\textwidth}
    		\centering
    		Data $\log T_1$
    	\end{subfigure}
    	\begin{subfigure}{0.49\textwidth}
    		\centering
    		Model $\log T_1$
    	\end{subfigure}	
    	\begin{subfigure}{0.49\textwidth}
    		\centering
    		\includegraphics[width = \textwidth]{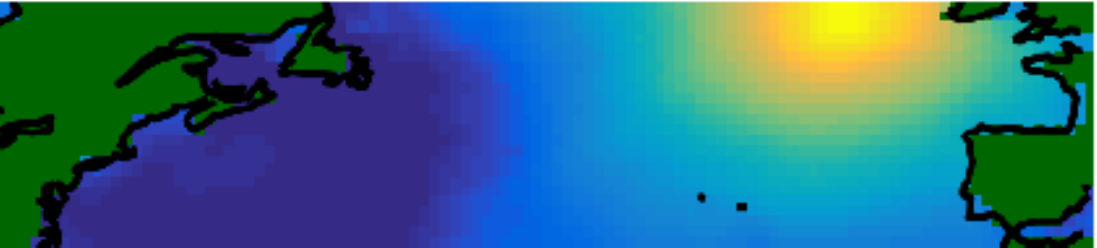}
    	\end{subfigure}
    	\begin{subfigure}{0.49\textwidth}
    		\includegraphics[width = \textwidth]{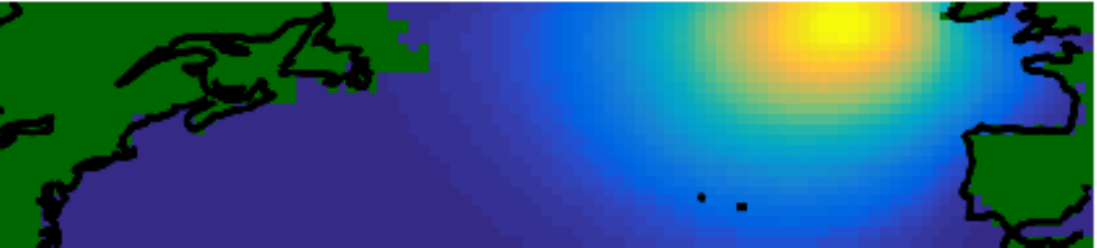}
    	\end{subfigure}
    	\begin{subfigure}{0.49\textwidth}
    		\centering
    		\includegraphics[width = \textwidth]{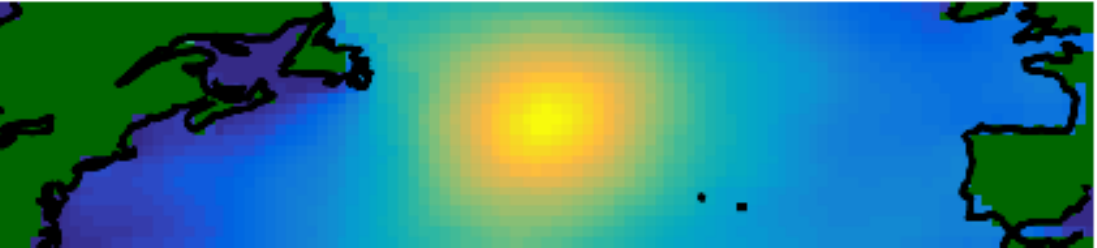}
    	\end{subfigure}
    	\begin{subfigure}{0.49\textwidth}
    		\includegraphics[width = \textwidth]{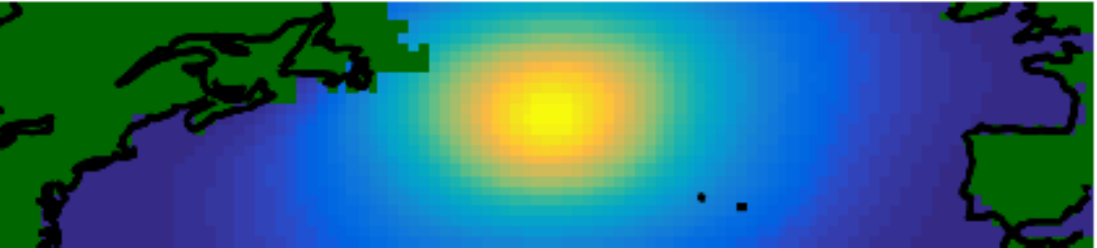}
    	\end{subfigure}
    	\begin{subfigure}{0.49\textwidth}
    		\centering
    		\includegraphics[width = \textwidth]{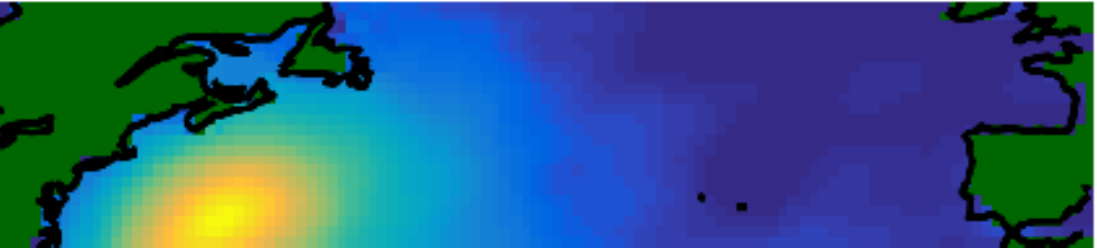}
    	\end{subfigure}
    	\begin{subfigure}{0.49\textwidth}
    		\includegraphics[width = \textwidth]{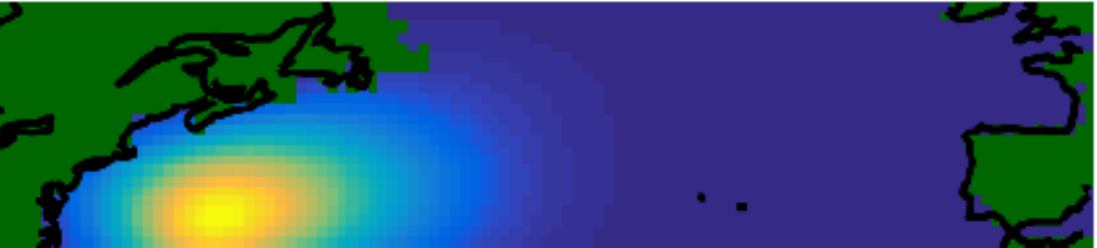}
    	\end{subfigure}	
	\end{subfigure}
	\begin{subfigure}{0.07\textwidth}
	\centering
	\includegraphics[width=\textwidth]{./figs/examples/colorbar01.png}
	\end{subfigure}	
	\caption{Correlation between three different reference points and all other points in $\log T_1$. Left column: empirical correlation function from data. Right column: correlation function from fitted model. }
	\label{fig:correlationTp}
\end{figure}

The estimated correlation functions for three reference points are visualized in Figures \ref{fig:correlationHs} and \ref{fig:correlationTp}. Thus, the figures show the correlation between the reference points and all other points in the domain.
These three reference points have the coordinates $296^{\circ}$ longitude, $37^{\circ}$ latitude (close to the east coast of USA), $320^{\circ}$ longitude, $44^{\circ}$ latitude (in the middle of the north Atlantic), and $342^{\circ}$ longitude, $51^{\circ}$ latitude (close to the west coast of Ireland).
The figures suggests that the correlation structures are quite similar between $\log H_s$ and $\log T_1$, which makes sense since they are positively correlated. 

The estimated smoothness parameter of $\log H_s$ was $\alpha = 3.66$, corresponding to a random field which is almost surely H{\"o}lder continuous with H{\"o}lder constant $2.66$. 
In \citet{lit:hildeman} the same model was fitted to $\log H_s$ with the difference that it was defined in the longitude-latitude projection instead of on the sphere and that the smoothness parameter could only be integer-valued. In that work, the smoothness was found to be $\alpha = 3$. With arbitrary smoothness we are now able to find a more exact estimate of the smoothness parameter. Likewise, the estimated smoothness of $\log T_1$ was $\alpha = 3.16$, corresponding to H{\"o}lder constant $2.16$. Hence, the wave period is spatially a little bit rougher compared to the significant wave height.


\subsection{Estimation of the cross-correlation structure by ML}
\begin{figure}[t]
	\centering
	\begin{subfigure}{0.9\textwidth}
	\centering
    	\begin{subfigure}{0.49\textwidth}
    		\centering
    		\includegraphics[width = \textwidth]{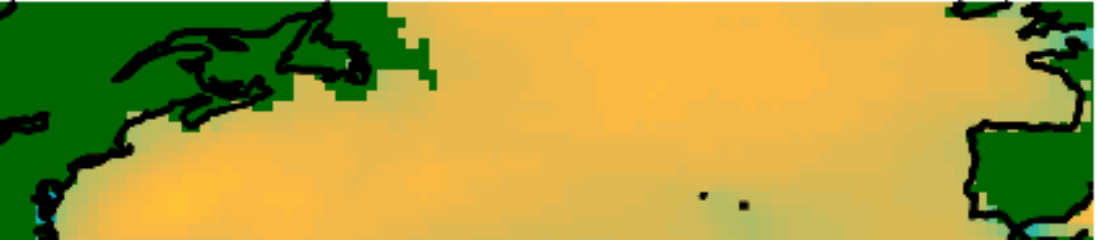}
    		\caption{Pointwise cross-correlation (data).}
    	\end{subfigure}	
    	\begin{subfigure}{0.49\textwidth}
    		\centering
    		\includegraphics[width = \textwidth]{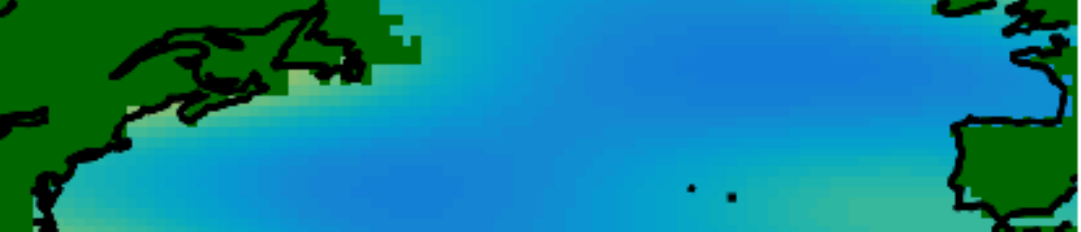}
    		\caption{Pointwise cross-correlation (model).}
    	\end{subfigure}\\
    	\begin{subfigure}{0.49\textwidth}
    		\centering
    		\includegraphics[width = \textwidth]{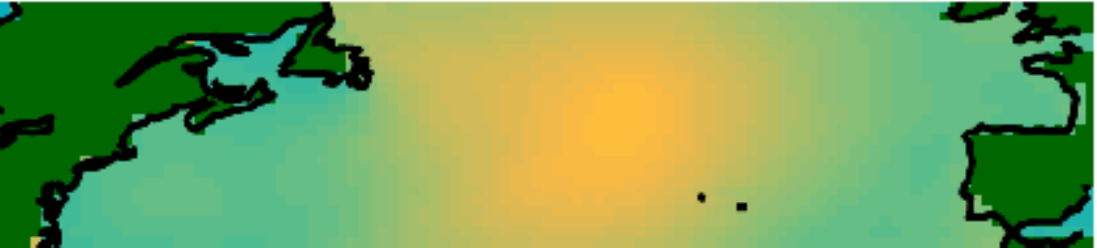}
    		\caption{Cross-correlation $T_1 \to H_s$ (data).}
    	\end{subfigure}
    	\begin{subfigure}{0.49\textwidth}
    		\centering
    		\includegraphics[width = \textwidth]{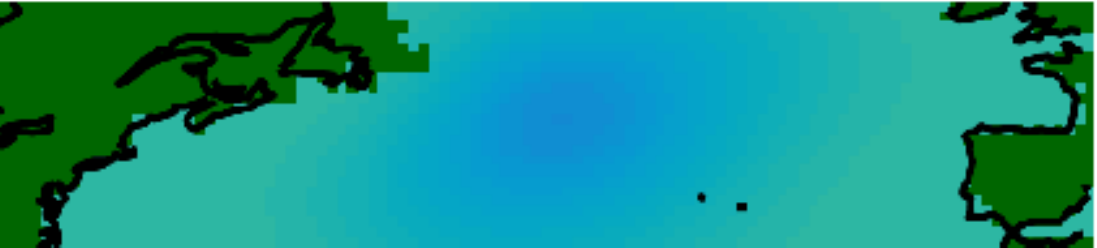}
    		\caption{Cross-correlation $T_1 \to H_s$ (model).}
    	\end{subfigure}
    	\begin{subfigure}{0.49\textwidth}
    		\centering
    		\includegraphics[width = \textwidth]{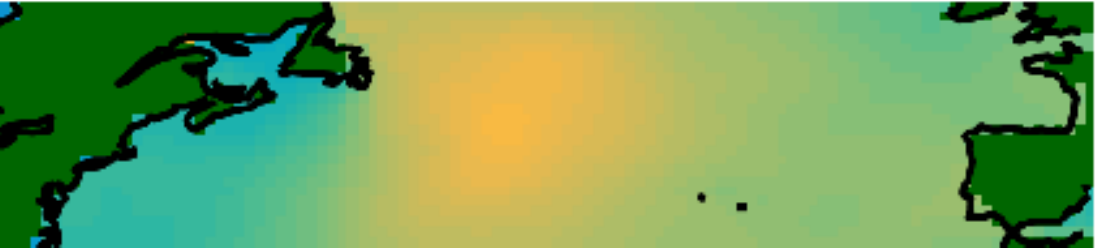}
    		\caption{Cross-correlation $H_s \to T_1$ (data).}
    	\end{subfigure}	
    	\begin{subfigure}{0.49\textwidth}
    		\centering
    		\includegraphics[width = \textwidth]{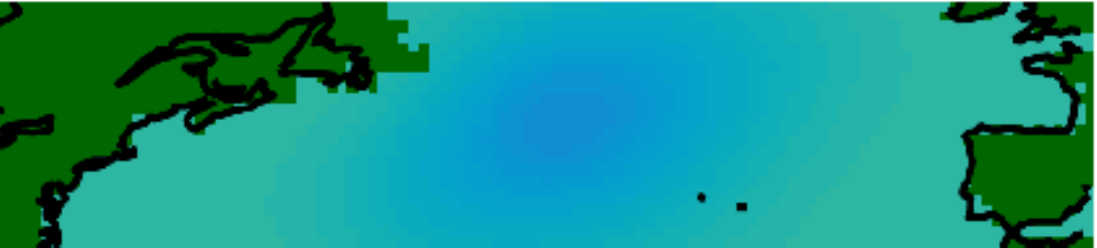}
    		\caption{Cross-correlation $H_s \to T_1$ (model).}
    	\end{subfigure}
    \end{subfigure}	
	\begin{subfigure}{0.07\textwidth}
	\centering
	\includegraphics[width= \textwidth]{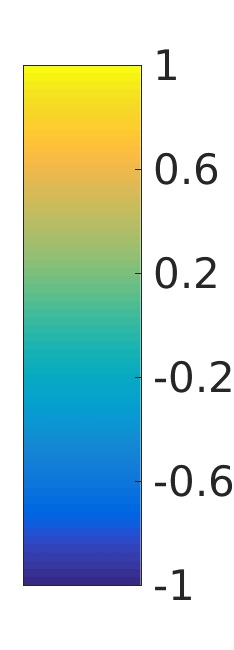}
	\end{subfigure}	
	\caption{Cross-correlation from joint maximum likelihood estimation of cross-correlation structure. Top row: comparison of, pointwise cross-correlation between the data and the estimated model. Middle row: comparison of cross-correlation between $\log T_1$ at a reference point and $\log H_s$ for all points in $\gspace$.
	Bottom row: comparison of cross-correlation between $\log H_s$ at a reference point and $\log T_1$ for all points in $\gspace$.}
	\label{fig:crosscorrelationProperRho}
\end{figure}

Given the marginal parameters of $X$ and $Y$, we now want to estimate their cross-correlation structure, i.e., $\rho(\psp)$. We parametrize this function as a regression on cosines as in \eqref{eq:cosineParams}.
Estimating $\rho(\psp)$ using ML conditioned on the already estimated parameters for $X$ and $Y$, we acquired parameters for our bivariate model of $H_s$ and $T_1$ jointly. Figure \ref{fig:crosscorrelationProperRho} compares the estimated cross-correlation structure with the empirical one estimated from data. The reference point used in this figure was at $320^{\circ}$ longitude and $44^{\circ}$ latitude.

\begin{figure}[t]
	\begin{center}
	\begin{subfigure}{0.4\textwidth}
		\centering
		\includegraphics[width = \textwidth]{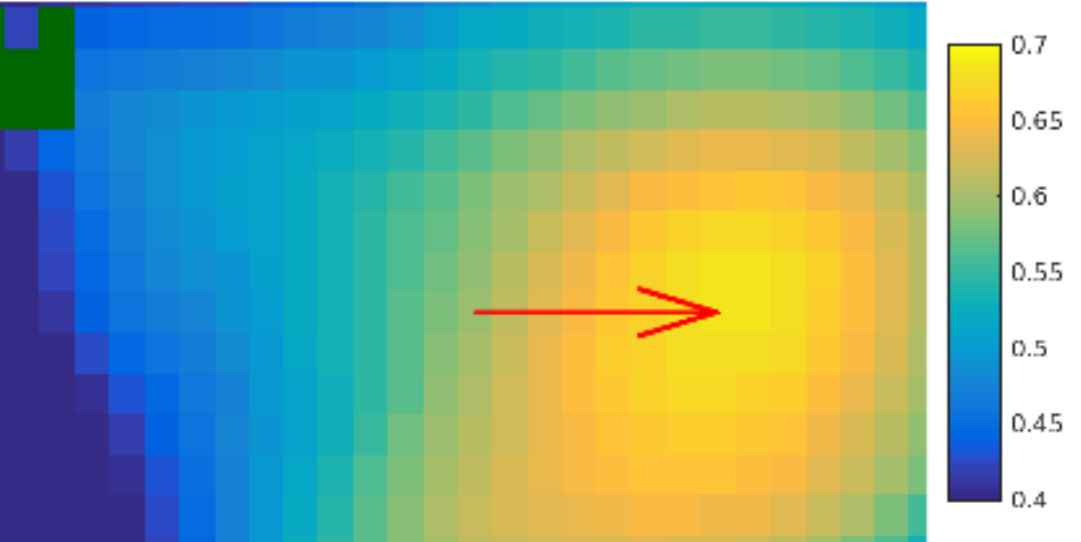}
		\caption{$T_1 \to H_s$}
	\end{subfigure}
	\begin{subfigure}{0.4\textwidth}
		\centering
		\includegraphics[width = \textwidth]{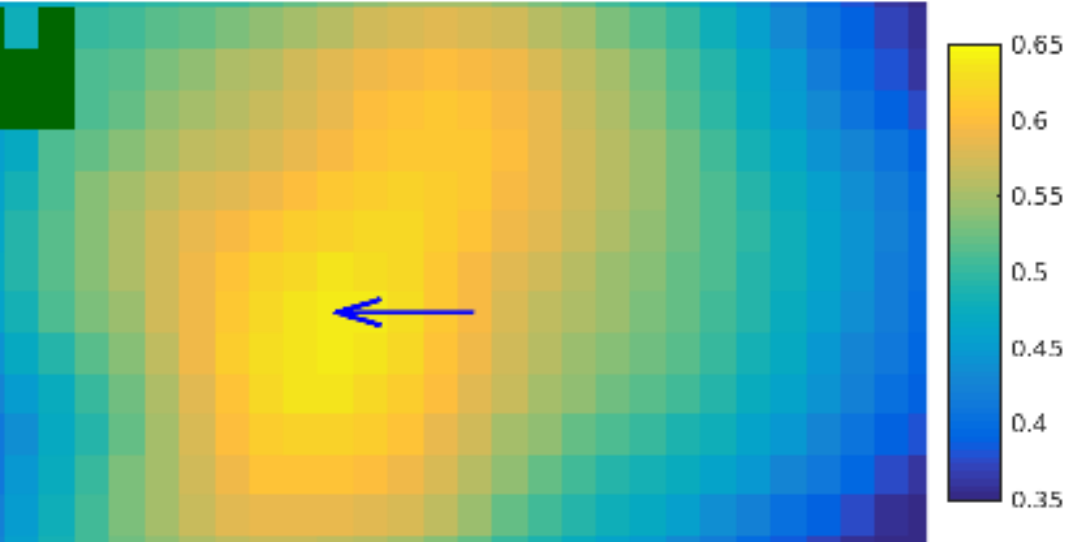}
		\caption{$H_s \to T_1$}
	\end{subfigure}\\
	\begin{subfigure}{0.6\textwidth}
		\centering
		\includegraphics[width = \textwidth]{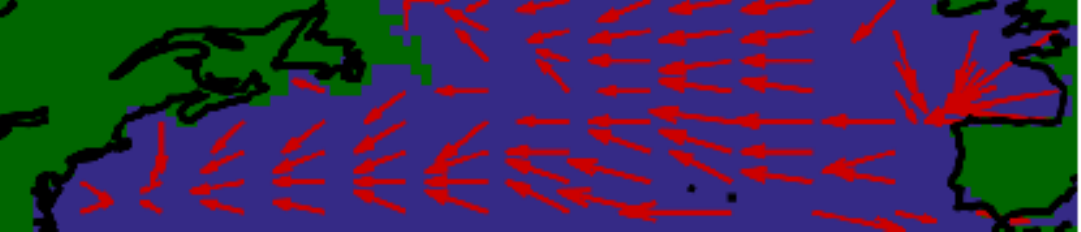}
		\caption{Shift of maximum cross-correlation.}
	\end{subfigure}
	\end{center}
	\vspace{-0.5cm}
	\caption{Top row: translation vectors between a reference point $\psp_r$ at $(320^{\circ}, 44^{\circ})$ and its maximum cross-correlation value (estimated from data). Cross-correlation between $\log T_1(\psp_r)$ and $\log H_s(\psp)$ for $\psp$ close to $\psp_r$ is shown to the left. The right panel show cross-correlation between $\log H_s(\psp_r)$ and $\log T_1(\psp)$.
	Bottom row: translation vectors between $H_s$ and corresponding maximum cross-correlation with $T_1$ for several points in the domain.}
	\label{fig:crosscorrelationTranslationRef}
\end{figure}

\begin{figure}[t]
\centering
\begin{subfigure}{0.6\textwidth}
\includegraphics[width = \textwidth]{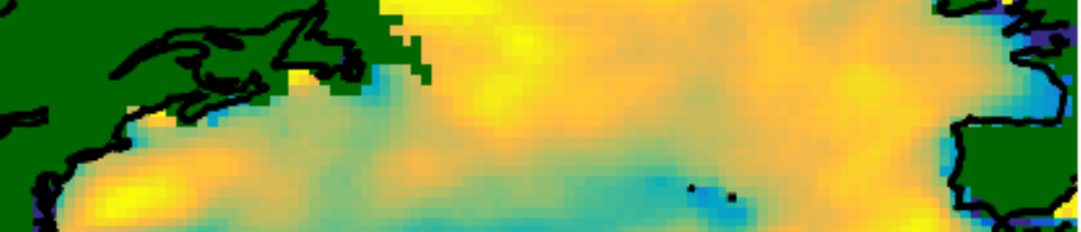}
\end{subfigure}
\begin{subfigure}{0.07\textwidth}
\includegraphics[width = 0.8 \textwidth]{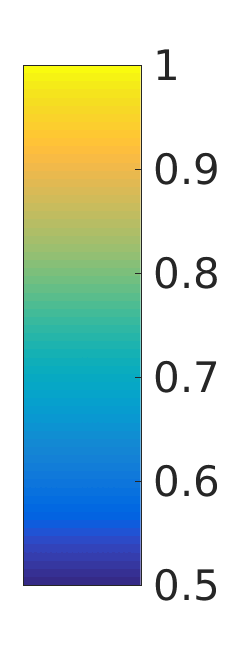}
\end{subfigure}
\caption{Ratio between maximum cross-correlation and pointwise cross-correlation for all points in the spatial domain. }
\label{fig:crosscorrelationTranslation}
\end{figure}

Surprisingly, even though the data is strongly positively correlated, the fitted model yielded a strong negative correlation. 
It turns out that the proposed model of the cross-correlation structure is a bit too simplistic to explain the true dependency between $\log H_s$ and $\log T_1$. The reason being that the point, $\psp_2$, where $\log T_1$ has the strongest cross-correlation with $\log H_s(\psp_1)$ is not $\psp_1$, i.e., $\psp_2 \neq \psp_1$. However, this is assumed in the proposed model of Section \ref{sec:model}.
For the reference point at $320^{\circ}$ longitude and $44^{\circ}$ latitude, the translation between the reference point and the point of maximum cross-correlation can be seen in Figure \ref{fig:crosscorrelationTranslationRef}. For $\log H_s$ in the reference point, corresponding $\log T_1$ is generally further west while the opposite relationship holds for $\log T_1$ in the reference point.
Corresponding vectors between reference points in $\log H_s$ and maximum points of correlation with $\log T_1$ can also be seen in the figure.
Figure \ref{fig:crosscorrelationTranslation} shows the ratio between the points of highest cross-correlation and the pointwise cross-correlation. 

For most regions, the pointwise cross-correlation is not that much smaller than the maximum cross-correlation. However, since there is a clear consistent increase in cross-correlation when moving away from the reference point, the maximum likelihood estimate of $\rho$ is negative. This obvious model-misspecification is another reason for using the proposed stepwise estimation procedure.

\subsection{Estimation of the cross-correlation structure by pointwise ML}
The results of the previous subsection suggest that the bivariate model will not explain the joint distribution perfectly. However, it can still be useful if one could obtain a better method of estimating $\rho$.
Instead of estimating $\rho$ by ML as before, a possible solution is to fit the model to explain the pointwise cross-correlation, intead of the total cross-correlation.
This corresponds to maximizing a product likelihood of the bivariate Gaussian random variables for each spatial location, i.e., the log-likelihood function 
\begin{align}
    l(\rho;  \bs{x}, \bs{y}) &= l(\rho; \bs{\hat{\gamma}}) = \sum_{j = 1}^{M} O_j \left[ -\log\left(2\pi\right) - \frac{1}{2}\log\left( 1-\gamma_j^2 \right)  
    +  \frac{\gamma_j}{1-\gamma_j^2} \left( \frac{O_j-1}{O_j} \hat{\gamma}_{j} \right)  \right].
    \label{eq:pointwiseML}
\end{align}
Here, $M$ is the number of locations where there have been observations in the data, $O_j$ are the number of observations for location $j$, and $\gamma_j$ is the pointwise cross-correlation between the two fields at location $\psp_j$ from the model. The observations, $\bs{x} := \{x_{jk}\}_{j,k}$ and $\bs{y} := \{y_{jk}\}_{j,k}$ are not needed explicitly since the sample pointwise cross-correlations, $\bs{\hat{\gamma}}$, are sufficient statistics for evaluating the log-likelihood.
The pointwise cross-correlations of the model are 
\begin{align}
\bs{\gamma} = A_{j\cdot} P_r \tilde{\Sigma}_{XY} Q_r^T A_{j\cdot}^T,
\end{align}
where $A$ is the $M\times N$ observational matrix, i.e., mapping nodal values to values at the locations of observations \citep{lit:lindgren}. The matrices $P_r$ and $Q_r$ are defined in Section \ref{sec:model} and are sparse $N\times N$ matrices. The matrices $\tilde{\Sigma}_{\star\star}$ are $N\times N$ block matrices of $\tilde{\Sigma}$ which is the covariance matrix of $[\tilde{U}_X, \tilde{U}_Y]$, as defined in Section \ref{sec:model}. 
To reduce the computational cost of computing $\Sigma_j$, we use the 
Takahashi equations \citep{lit:takahashi, lit:rue2} to compute the needed elements of $\tilde{\Sigma}$ based on the corresponding precision matrix---without computing the full inverse which is non-sparse. 

\begin{figure}[t]
	\centering
	\begin{subfigure}{0.44\textwidth}
		\centering
		\includegraphics[width = \textwidth]{./figs/estimation/crosscorrelation/data/crosscorFieldData.eps}
		\caption{Data.}
	\end{subfigure}
	\begin{subfigure}{0.44\textwidth}
		\includegraphics[width = \textwidth]{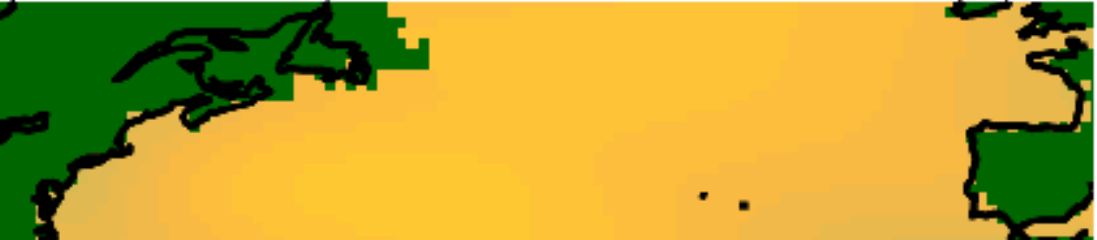}
		\caption{Model with pointwise ML estimates.}
	\end{subfigure}
	\begin{subfigure}{0.07\textwidth}
	\centering
	\raisebox{0.2cm}{\includegraphics[ width= 0.8 \textwidth]{./figs/examples/colorbarm11.png}}
	\end{subfigure}
	\vspace{-0.4cm}
	\caption{Comparison between the pointwise cross-correlation of data and the model using the pointwise ML estimates.}
	\label{fig:crosscorrelationDataField}
\end{figure}

When we estimated the parameters, the pointwise sample cross-correlations, $\{\hat{\gamma}_j\}_j$ were replaced with the sample cross-correlations between $H_s$ at location $\psp_j$ and $T$ at the location which maximized the pointwise cross-correlation. In this way, the fitted model will have a pointwise cross-correlation corresponding to the maximum cross-correlation of that point---instead of fitting a perfect pointwise cross-correlation that will underestimate the maximum cross-correlation somewhat. 
The pointwise cross-correlation as compared to data can be seen in Figure \ref{fig:crosscorrelationDataField}. As seen, the model has a larger pointwise cross-correlation, as designed.  

\begin{figure}[t]
	\centering
	\begin{subfigure}{0.9\textwidth}
	\centering
    	\begin{subfigure}{0.49\textwidth}
    		\centering
    		Data $T_1 \to H_s$
    	\end{subfigure}
    	\begin{subfigure}{0.49\textwidth}
    		\centering
    		Model $T_1 \to H_s$
    	\end{subfigure}	
    	\begin{subfigure}{0.49\textwidth}
    		\centering
    		\includegraphics[width = \textwidth]{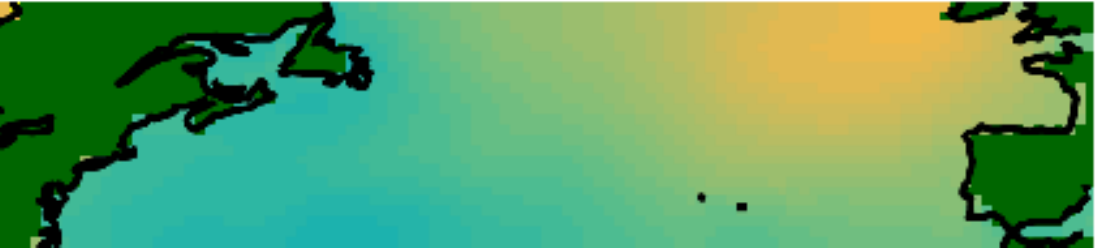}
    	\end{subfigure}
    	\begin{subfigure}{0.49\textwidth}
    		\includegraphics[width = \textwidth]{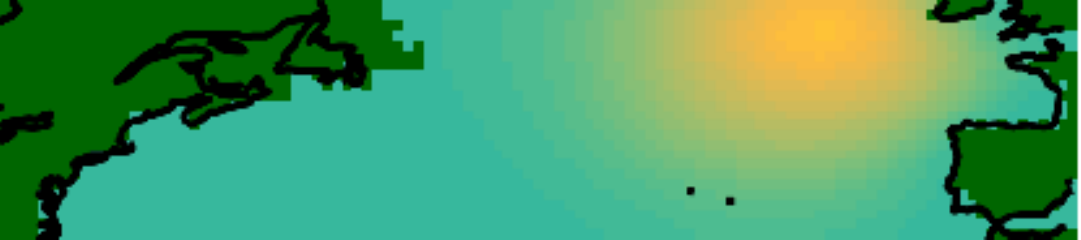}
    	\end{subfigure}
    	\begin{subfigure}{0.49\textwidth}
    		\centering
    		\includegraphics[width = \textwidth]{./figs/estimation/crosscorrelation/data/atlantic/crosscorPointHsTpData.eps}
    	\end{subfigure}
    	\begin{subfigure}{0.49\textwidth}
    		\includegraphics[width = \textwidth]{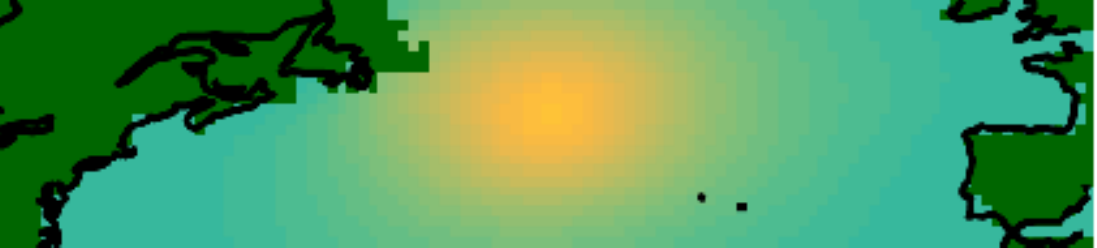}
    	\end{subfigure}
    	\begin{subfigure}{0.49\textwidth}
    		\centering
    		\includegraphics[width = \textwidth]{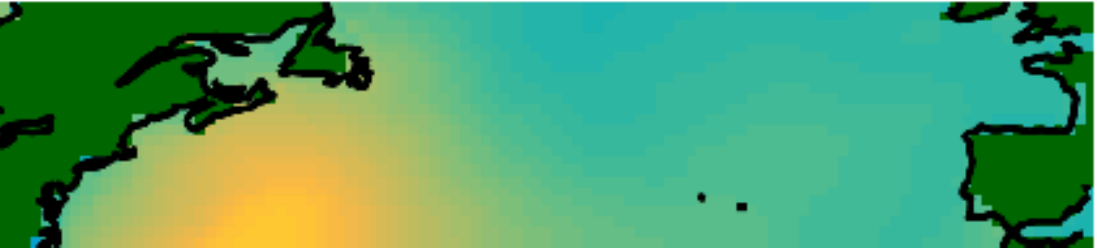}
    	\end{subfigure}
    	\begin{subfigure}{0.49\textwidth}
    		\includegraphics[width = \textwidth]{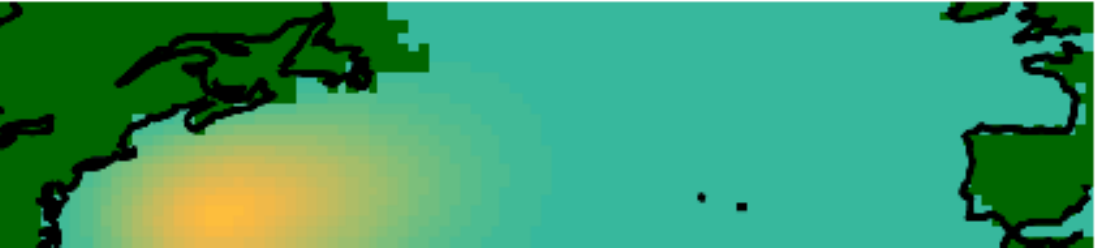}
    	\end{subfigure}	
	\end{subfigure}
	\begin{subfigure}{0.08\textwidth}
	\centering
	\includegraphics[width= 0.8 \textwidth]{./figs/examples/colorbarm11.png}
	\end{subfigure}	
	\caption{Cross-correlation between three reference points in $\log T_1$ and all other points in $\log H_s$. Left column: empirical cross-correlation function from data. Right column: cross-correlation function from fitted model using pointwise ML estimates.}
	\label{fig:crosscorrelationHsTpMaxRho}
\end{figure}

\begin{figure}[t]
	\centering
	\begin{subfigure}{0.9\textwidth}
	\centering
    	\begin{subfigure}{0.49\textwidth}
    		\centering
    		Data $H_s \to T_1$
    	\end{subfigure}
    	\begin{subfigure}{0.49\textwidth}
    		\centering
    		Model $H_s \to T_1$
    	\end{subfigure}	
    	\begin{subfigure}{0.49\textwidth}
    		\centering
    		\includegraphics[width = \textwidth]{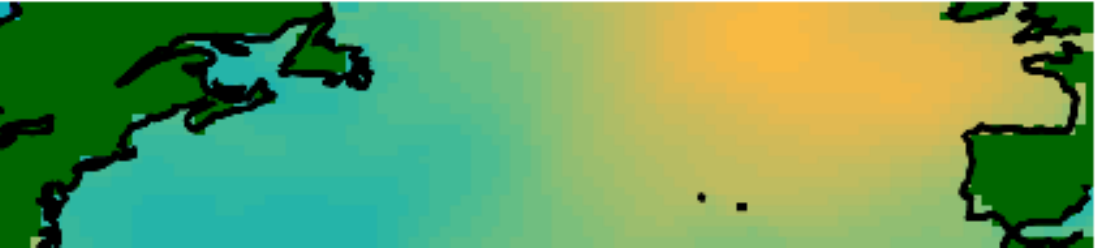}
    	\end{subfigure}
    	\begin{subfigure}{0.49\textwidth}
    		\includegraphics[width = \textwidth]{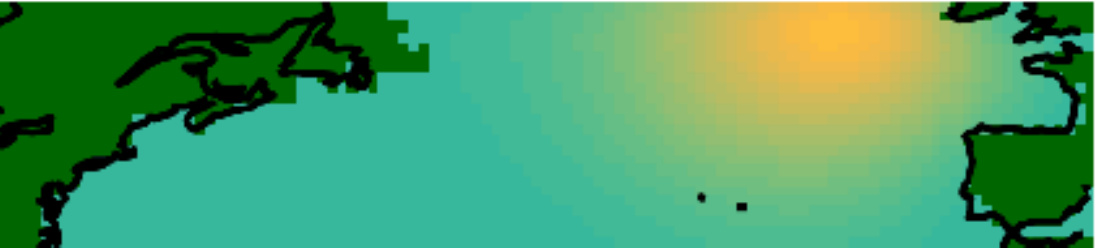}
    	\end{subfigure}
    	\begin{subfigure}{0.49\textwidth}
    		\centering
    		\includegraphics[width = \textwidth]{./figs/estimation/crosscorrelation/data/atlantic/crosscorPointTpHsData.eps}
    	\end{subfigure}
    	\begin{subfigure}{0.49\textwidth}
    		\includegraphics[width = \textwidth]{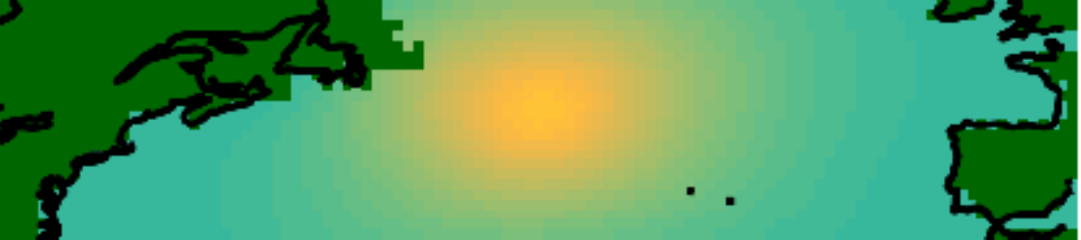}
    	\end{subfigure}
    	\begin{subfigure}{0.49\textwidth}
    		\centering
    		\includegraphics[width = \textwidth]{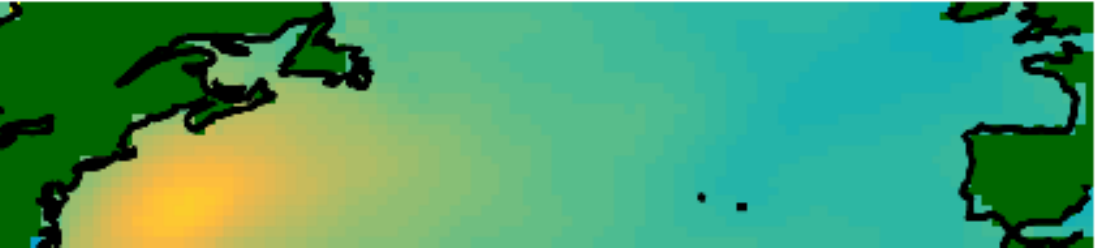}
    	\end{subfigure}
    	\begin{subfigure}{0.49\textwidth}
    		\includegraphics[width = \textwidth]{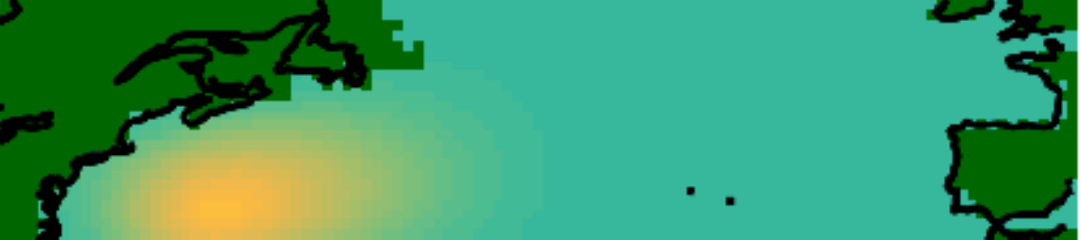}
    	\end{subfigure}	
	\end{subfigure}
	\begin{subfigure}{0.08\textwidth}
	\centering
	\includegraphics[width= 0.8 \textwidth]{./figs/examples/colorbarm11.png}
	\end{subfigure}
\caption{Cross-correlation between three reference points in $\log H_s$ and all other points in $\log T_1$. Left column: empirical cross-correlation function from data. Right column: cross-correlation function from fitted model using pointwise ML estimates.}
	\label{fig:crosscorrelationTpHsMaxRho}
\end{figure}

\begin{figure}[t]
	\centering
    \begin{subfigure}{0.49\textwidth}
	\centering		
	\includegraphics[width=\textwidth]{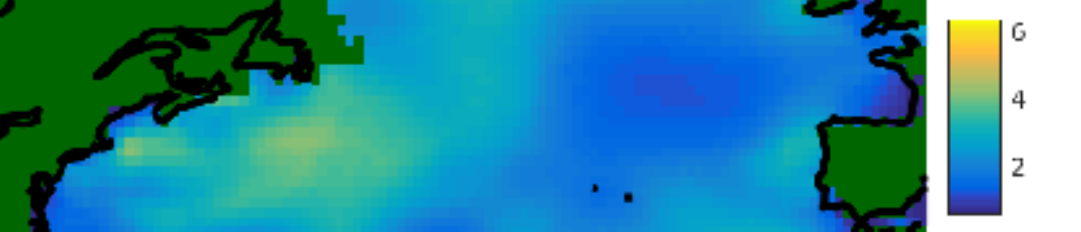}
	\caption{$H_s$ simulation $1$}		
	\end{subfigure}
	\begin{subfigure}{0.49\textwidth}
	\centering		
	\includegraphics[width=\textwidth]{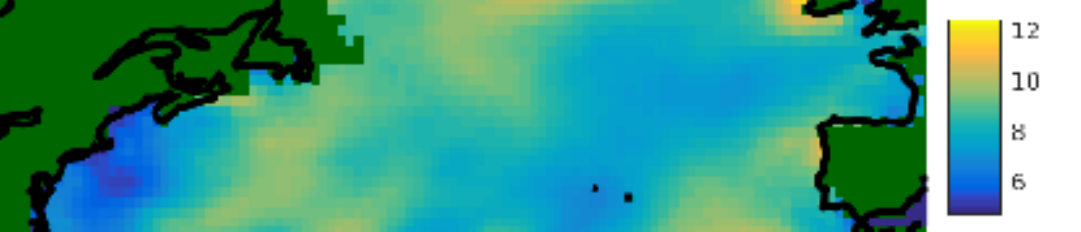}	
	\caption{$T_1$ simulation $1$}		
	\end{subfigure}		\\
    \begin{subfigure}{0.49\textwidth}
	\centering		
	\includegraphics[width=\textwidth]{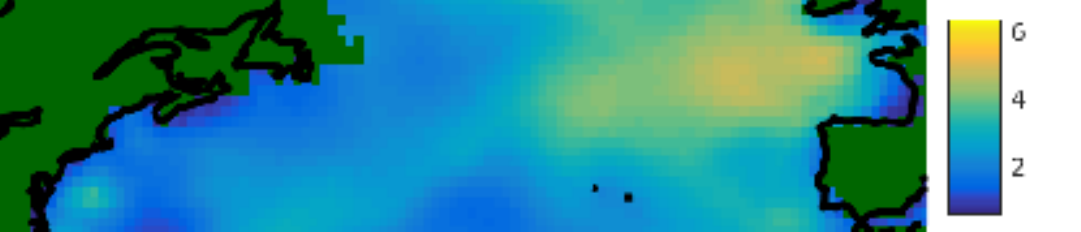}
	\caption{$H_s$ simulation $2$}		
	\end{subfigure}
	\begin{subfigure}{0.49\textwidth}
	\centering		
	\includegraphics[width=\textwidth]{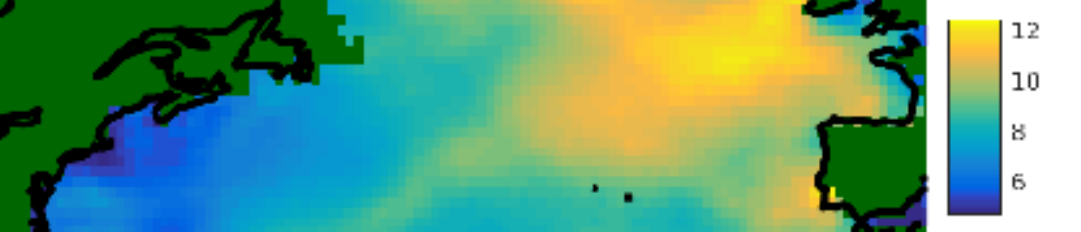}	
	\caption{$T_1$ simulation $2$}		
	\end{subfigure}			
	\caption{Two simulations of $H_s$ and $T_1$ from the joint spatial model.}
	\label{fig:realizations2}
\end{figure}

To get an understanding of the true cross-correlation structure of the estimated parameters, Figures \ref{fig:crosscorrelationHsTpMaxRho} and \ref{fig:crosscorrelationTpHsMaxRho} show the cross-correlation between the three reference points in one of the fields and all points in the other field. 
Finally, Figure~\ref{fig:realizations2} shows realizations from the final model, which look similar to the observed data in Figure \ref{fig:realizations}.

\section{Applications}
\label{sec:applications}
In this section we look into two applications in maritime safety for which information about both $H_s$ and $T$ are used. One is an extension of the fatigue damage application considered in \citet{lit:hildeman}. The other is a method of estimating the risk of capsizing due to a specific capsizing mode known as \textit{broaching-to}.

\begin{figure}[t]
	\centering
	\begin{subfigure}{0.4\textwidth}
	    \centering
    	\includegraphics[width= 0.8\textwidth]{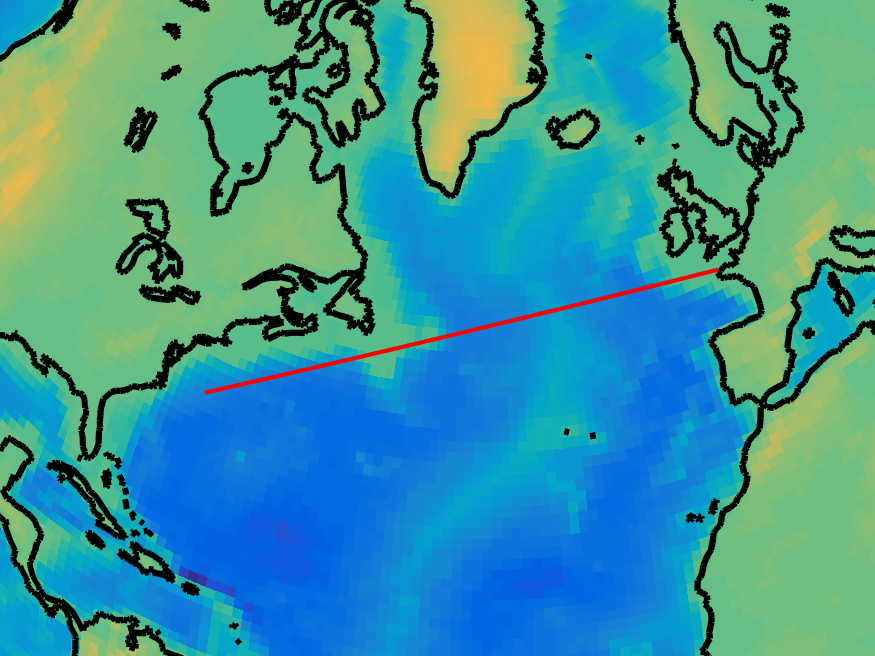}
    	\caption{Transatlantic route of ship.}
    	\label{fig:route}
	\end{subfigure}
	\begin{subfigure}{0.4\textwidth}
	    \centering
    	\includegraphics[width= 0.8\textwidth]{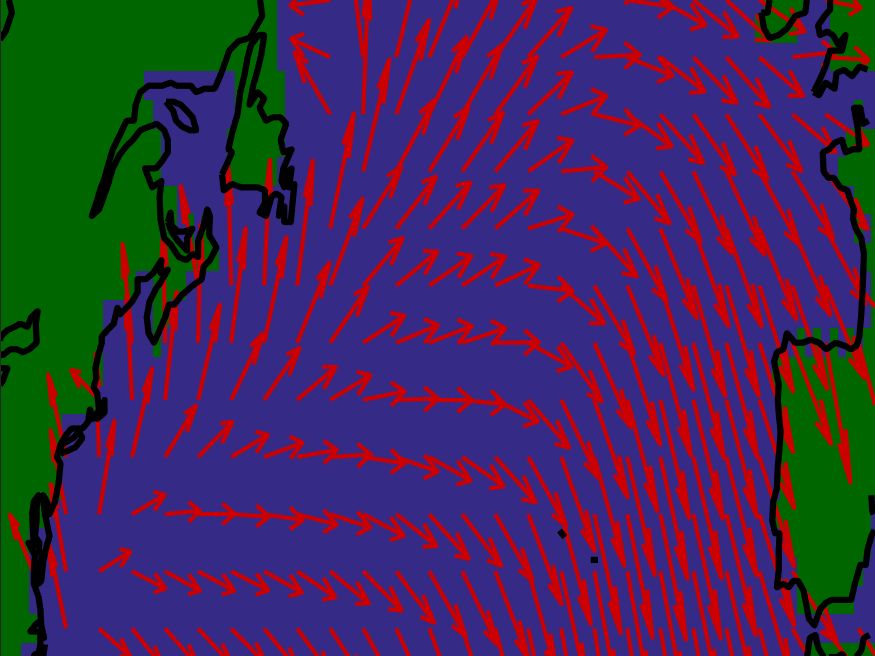}
    	\caption{Mean wave directions during April.}
    	\label{fig:waveDirection}
	\end{subfigure}
	\caption{Route of ship and the mean wave directions during month of April. }
\end{figure}


\subsection{Accumulated fatigue damage}
\label{sec:fatigue}
A ship traversing the ocean is subjected to wear due to collisions with waves. These collisions will create microscopic cracks in the hull of the ship. With time and further exposure to the wave environment such cracks will grow while new will form.  This type of wear damage is called fatigue. A ship will accumulate a certain amount of fatigue damage on any journey. However, the accumulated fatigue damage will vary in severity depending on the sea states encountered en route.
\citet{lit:mao} proposed the following formula based on $H_s$ and $T_z$ for which the expected rate, $d(t)$, of accumulated fatigue damage could be computed, 
\begin{align}
d(t) \approx
\frac{0.47 C^{\beta} H_s^\beta(\psp(t))}{\gamma} \left( \frac{1}{T_z(\psp(t))} - \frac{2\pi V(t)\cos \alpha(t)}{gT_z^2(\psp(t))} \right).
\label{eq:fatigueDamage}
\end{align}
Here, $g$ is the gravitational constant ($\approx 9.81$), $V$ is the speed of the ship, 
and $\alpha$ is the angle between the heading of the ship and the direction of the traveling waves. Further, $\gamma$ and $\beta$ are constants dependent on the material of the ship and $C$ is a constant depending on the ship's design \citep{lit:mao}. This formula can be used in combination with Monte Carlo simulations of $H_s$ and $T_z$ from our proposed model to evaluate the distribution of accumulated fatigue damage on a planned route. 


We consider the transatlantic route of Figure \ref{fig:route}. The continuous route is approximated by line segments between $100$ point locations (evenly spaced in geodesic distance). We set the ship speed to a fixed value of $10$ [m/s] which yields a sailing duration of $149.69$ hours or equivalently $6.23$ days. The heading of the ship, in one of the 100 locations on the route, is approximated as the mean between the direction acquired from the two connecting line segments. We consider the journey to take place in April, since we have estimated the parameters of the model for this month. A ship traversing the considered route can be modeled by a curve in space and time, $\psp_{\gamma}(t)$. Since we have neither a spatio-temporal model nor data with sufficient temporal resolution, we consider the sea states remaining constant in time during the traversal of the route, i.e., $\psp_{\gamma}(t) \in \gspace$ and not in space-time. 
We denote the accumulated fatigue damage during the trip up until time $t$ as $D(t)$, where $t = 0$ corresponds to the start of the trip, with no accumulated damage, and $t = t_{end}$ corresponds to the end of the trip, with maximal accumulated damage. 
We set the constants specific to the ship as in \citep{lit:podgorski, lit:mao, lit:hildeman}, i.e., $C = 20, \beta = 3,$ and $\gamma = 10^{12.73}$. 
In order to compute the fatigue, we also need the propagating waves angle in comparison with the ships heading. This is a random quantity that is not modeled in this work. Instead we assume that the mean direction of the wave propagation is the same as the direction that the countour lines of $H_s$ moves, i.e. the direction of the gradient of $H_s$ field (this is the same wave direction as used in \citep{lit:podgorski, lit:mao, lit:hildeman}, which has shown good results). 
This direction was estimated in \citet{lit:baxevani} and can be seen in Figure \ref{fig:waveDirection}.
Furthermore, assuming that the sea states can be characterized by Bretschenider spectrums, $T_z = \frac{1.2965}{1.408} T_1 = 0.9208 \cdot T_1$.
With these assumptions, and given values of $H_s$ and $T_1$, we use \eqref{eq:fatigueDamage} to compute the corresponding values of $d(t)$, and approximate the accumulated fatigue damage as
\begin{align}
D(T) = \int_0^{t_{end}} d(t)dt \approx \sum_{i=1}^{100} d(t_i) \Delta t,
\end{align}
where $\Delta t$ is the time differences between the $100$ consecutive point locations on the considered route, $\Delta t = 1.4969 $ hours.

\begin{figure}[t]
	\centering
	\begin{subfigure}{0.4\textwidth}
		\centering
		\includegraphics[width = \textwidth]{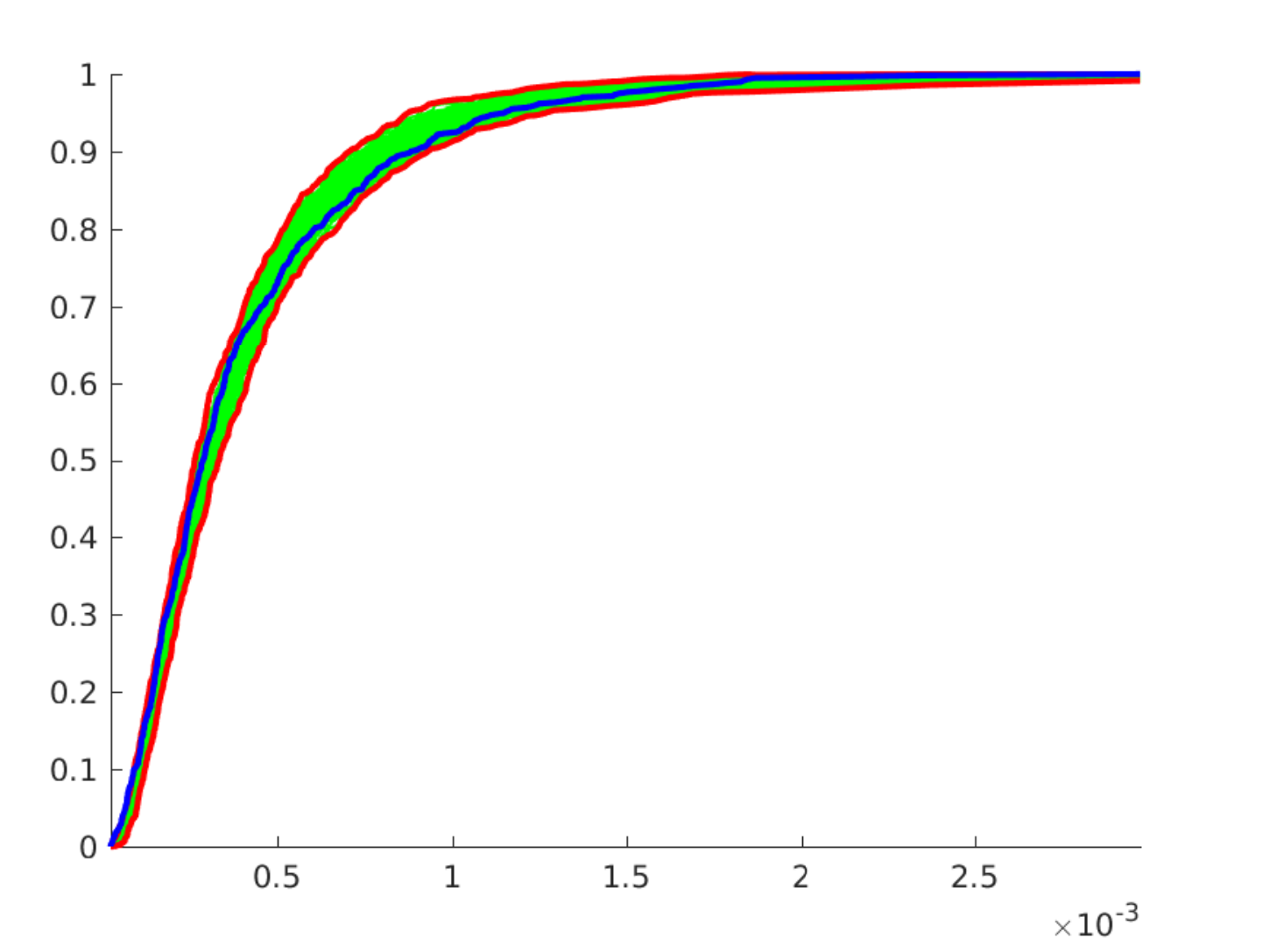}
		\caption{From USA to Europe.}
	\end{subfigure}
	\begin{subfigure}{0.4\textwidth}
		\centering
		\includegraphics[width = \textwidth]{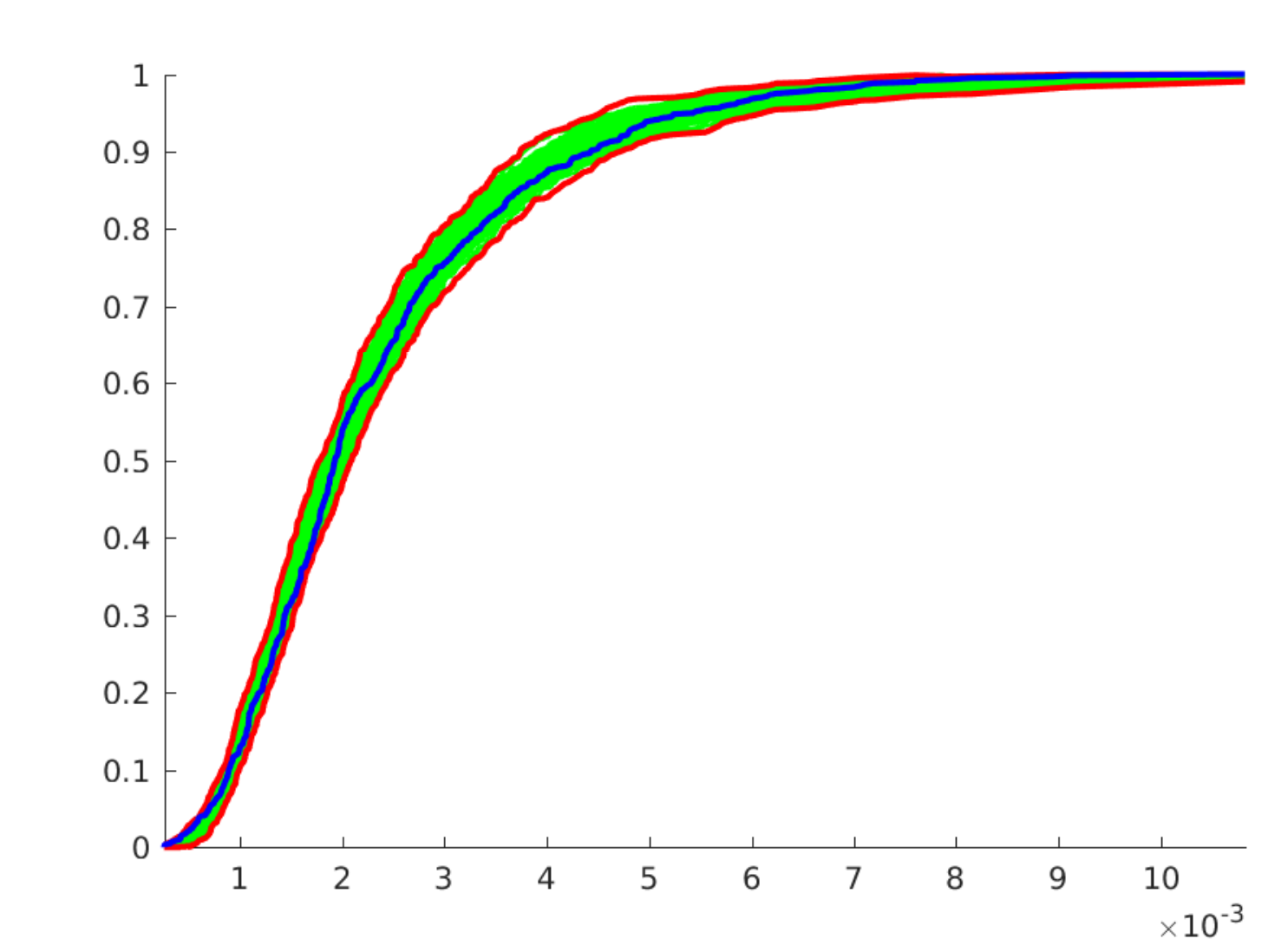}
		\caption{From Europe to USA.}
	\end{subfigure}
	\caption{Empirical CDFs of accumulated fatigue damage for the transatlantic route. Empirical CDF from data (blue line), $200$ different empirical CDFs from simulations (green), and pointwise upper and lower envelopes of the simulated CDFs (red). }
	\label{fig:fatigueMaximumRho}
\end{figure}

The accumulated fatigue damage is computed for each of the $600$ days available in the test set of the data. Hence, we acquire a sample of $600$ values of accumulated fatigue damage. Figure \ref{fig:fatigueMaximumRho} shows the empirical CDF computed from this sample (blue line). The accumulated damage is computed for a ship traversing the route in both directions, since the accumulated damage will depend on the angle between the heading of the ship and the propagation direction of the waves. 
In order to assess whether the estimated CDF from data behaves as if estimated from the model, we also estimate 200 CDFs from independent sets of data generated from the model. That is, 200 times we generate 600 independent realizations of the bivariate $H_s, T$ surface, and from each set of 600 realizations we compute a CDF. In the figure, these 200 estimated CDFs are plotted (green lines) together with the pointwise upper and lower envelopes of the values (red lines).
As can be seen, the estimated CDF from data is within the envelopes, suggesting that the model can be used for fatigue damage predictions.

%

\begin{figure}[t]
	\centering
	\begin{subfigure}{0.4\textwidth}
		\centering
		\includegraphics[width = \textwidth]{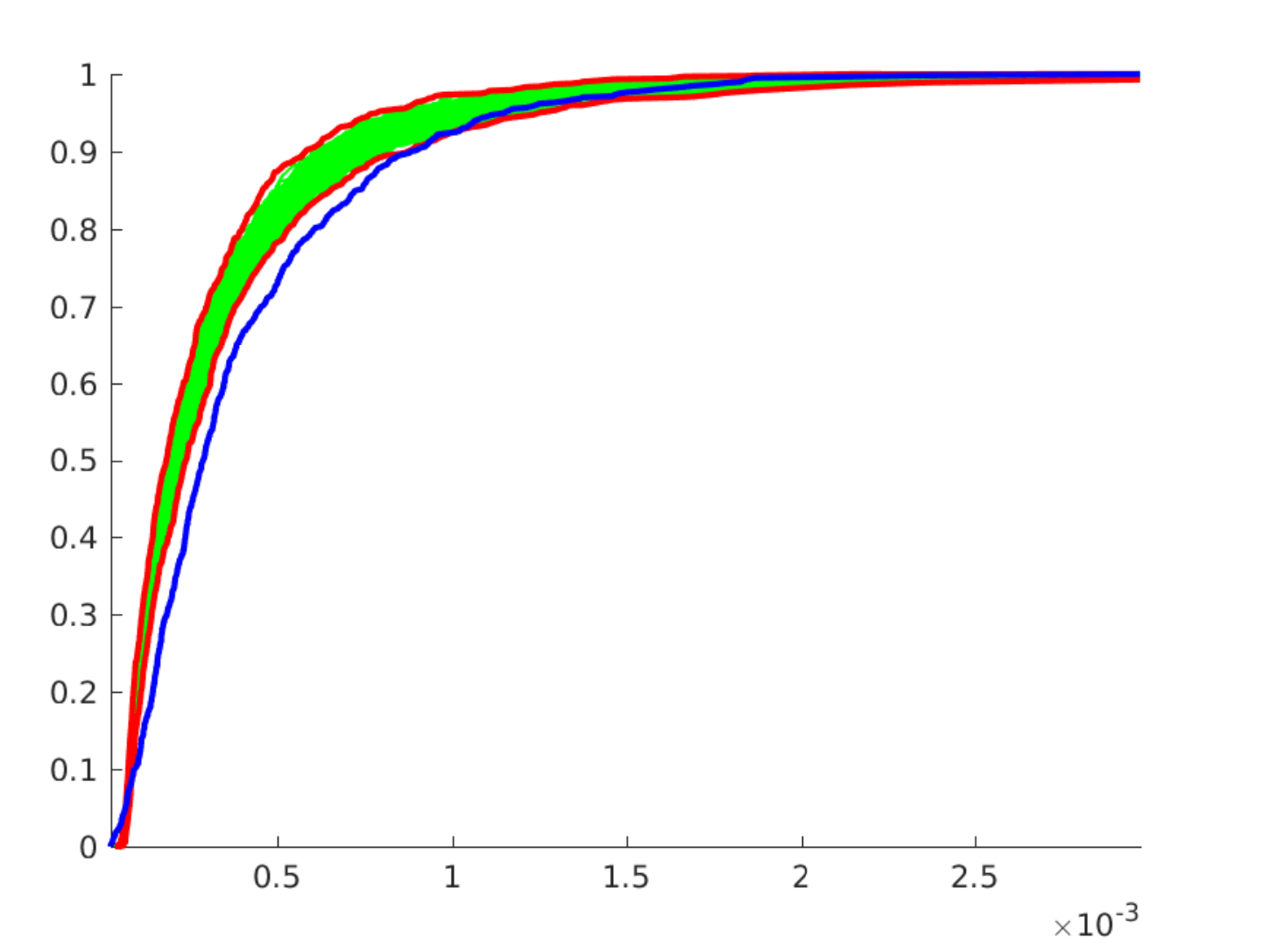}
		\caption{From USA to Europe.}
	\end{subfigure}
	\begin{subfigure}{0.4\textwidth}
		\centering
		\includegraphics[width = \textwidth]{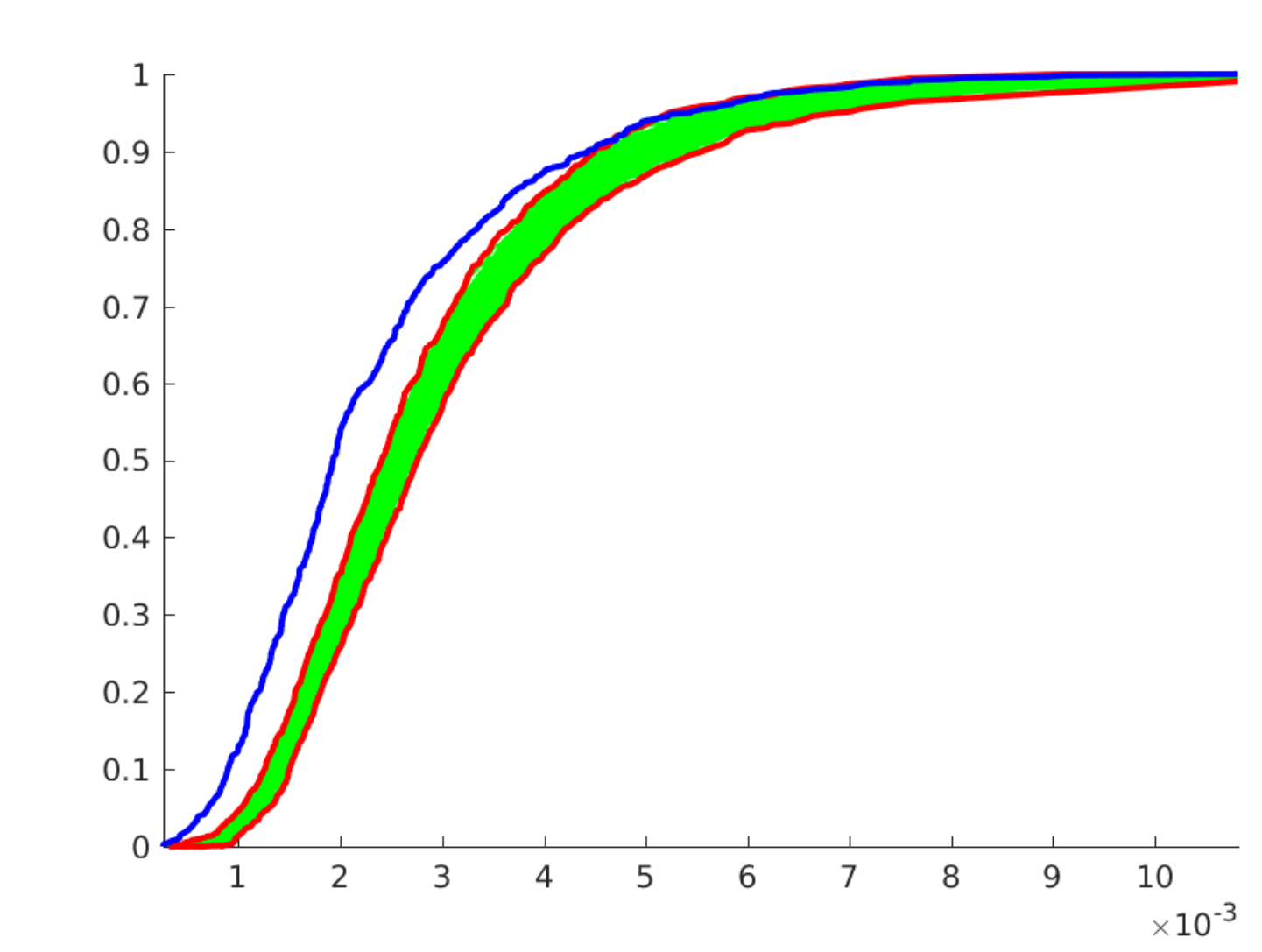}
		\caption{From Europe to USA.}
	\end{subfigure}
	\caption{Results for accumulated fatigue damage as in Figure \ref{fig:fatigueMaximumRho} but where the univariate spatial model of $H_s$ was used together with the proxy $T_z(\psp) = 3.75\sqrt{H_s(\psp)}$.}
	\label{fig:fatiguePureHs}
\end{figure}

In \citet{lit:hildeman}, a similar comparison was performed where the accumulated fatigue damage was computed using only $H_s$. Instead of $T_z$, the proxy $T_z = 3.75 \sqrt{H_s}$ was used, as proposed in~\citep{lit:mao, lit:podgorski}.
\citet{lit:hildeman} showed that the accumulated fatigue damage of the model agreed well with observed data. However, in that work only data of $H_s$ was available. Hence, the data that the model was compared to also used the proxy $T_z = 3.75 \sqrt{H_s}$. Since we have data of both $H_s$ and $T$, we can compare this proxy with data from the real bivariate random field. Figure \ref{fig:fatiguePureHs} shows the corresponding CDFs, and one can note that the use of the proxy does not provide accurate estimates of the true distribution of fatigue damage. In the direction from America to Europe, the model underestimates the damage, while in the other direction it overestimates it. This suggests that it is necessary to use a bivariate model in order to model accumulated fatigue damage correctly. 

\begin{figure}[t]
	\centering
	\begin{subfigure}{0.4\textwidth}
		\centering
		\includegraphics[width = \textwidth]{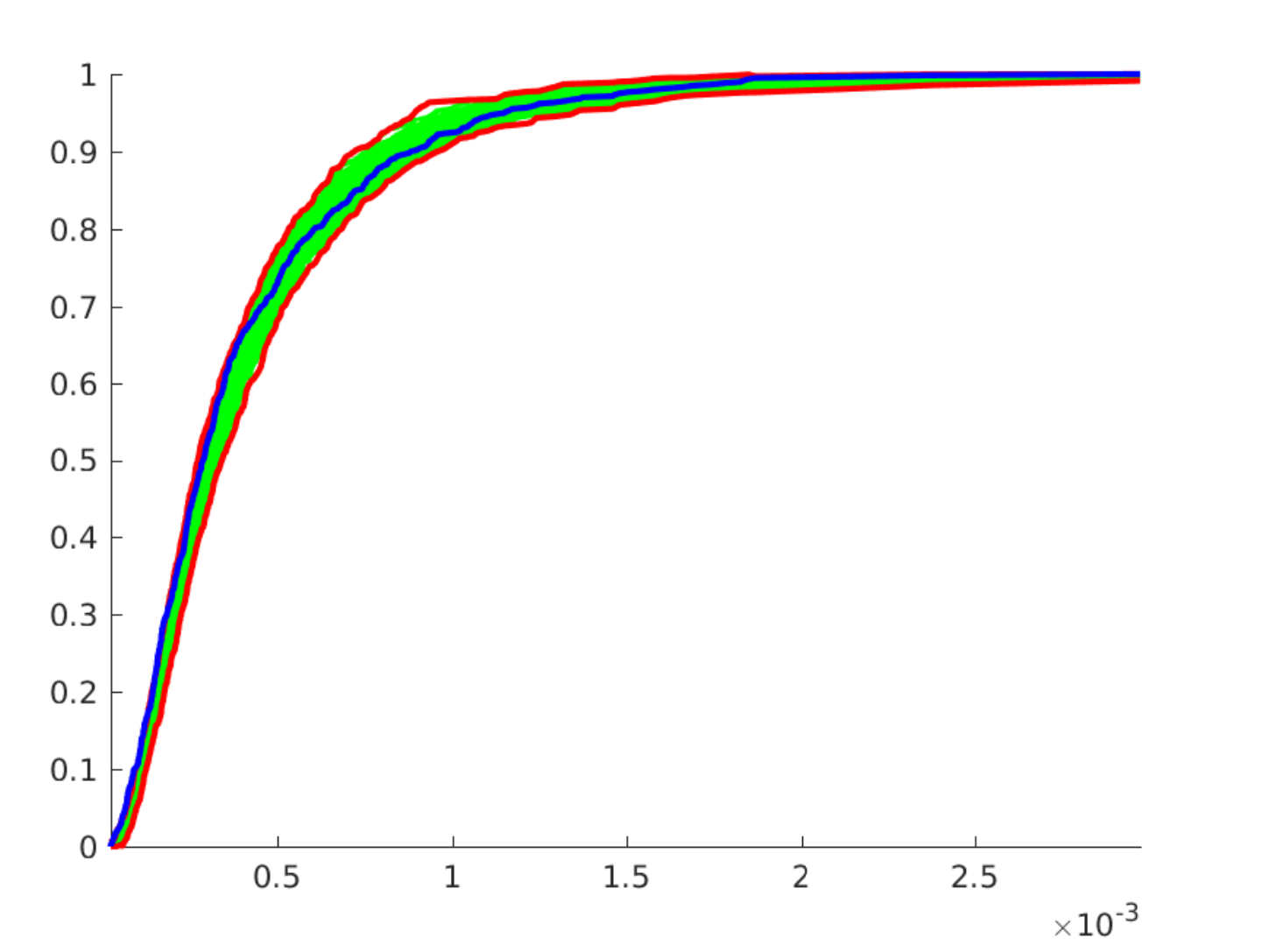}
		\caption{From USA to Europe.}
	\end{subfigure}
	\begin{subfigure}{0.4\textwidth}
		\centering
		\includegraphics[width = \textwidth]{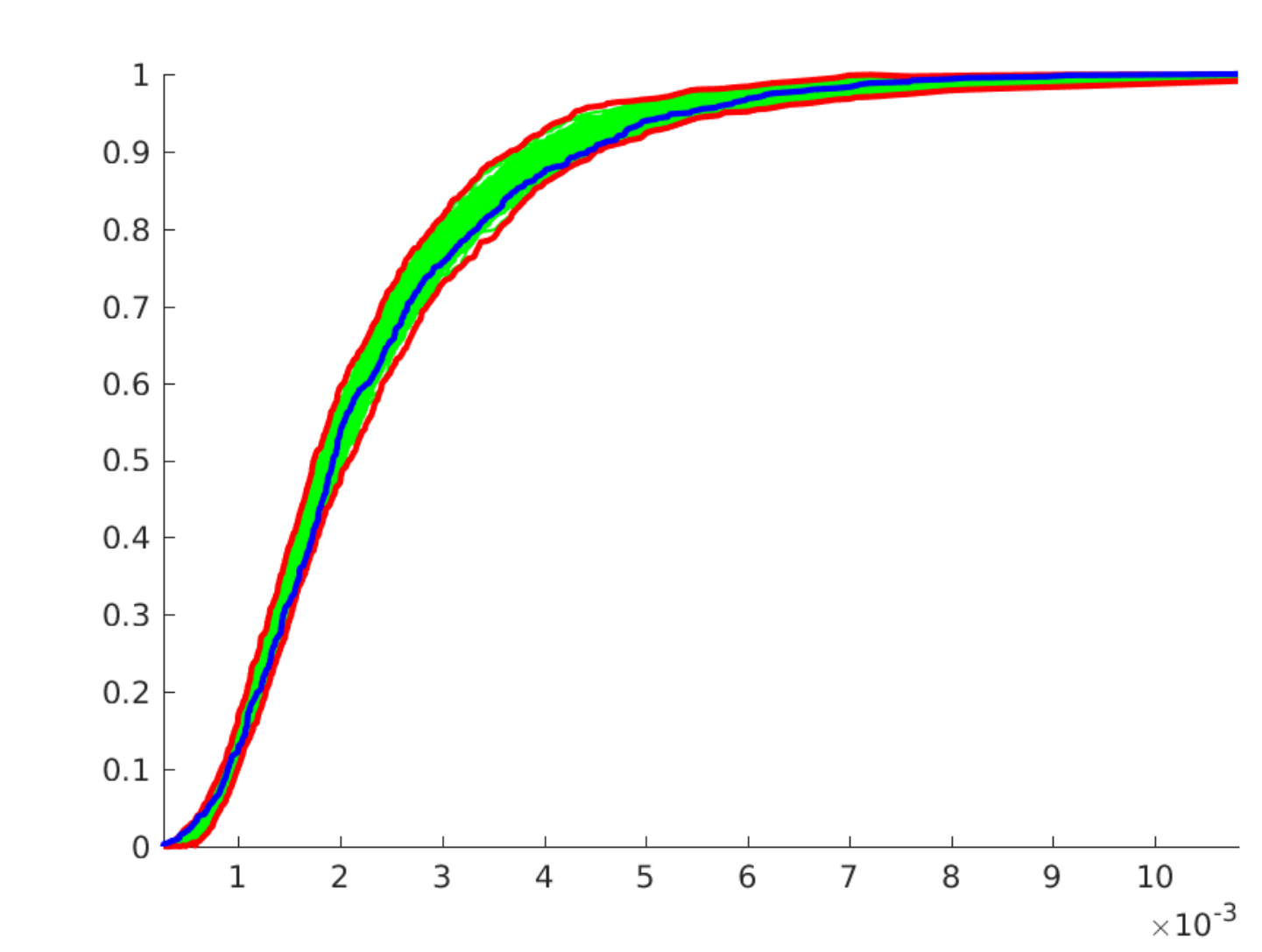}
		\caption{From Europe to USA.}
	\end{subfigure}
	\caption{Results for accumulated fatigue damage as in Figure \ref{fig:fatigueMaximumRho} but where the univariate spatial model of $H_s$ is used and $T_z$ is the conditional mean given $H_s$ (when using the pointwise ML estimate).}
	\label{fig:fatigueMaxRhoPureHsConditionalMean}
\end{figure}

However, instead of using the full bivariate model, a possible simpler alternative is to model $T$ as 
the pointwise conditional mean given $H_s$. In such a model, only $H_s$ has to be modeled spatially. Compared to the proxy model of \citet{lit:hildeman}, the pointwise cross-correlation between $H_s$ and $T$ would still need to be estimated.
Using this conditional means model for $T_z$ given $H_s$ yields the estimated CDFs as in Figure \ref{fig:fatigueMaxRhoPureHsConditionalMean}.
Also, this simpler model seemed sufficient to explain the distribution of fatigue damage accumulated on the transatlantic route. 


\subsection{Safety of operation in a following sea}
\label{sec:broaching}
Although capsizing of ships is rare, it is an important issue in naval architecture of hull designs of new vessels as well as for operational recommendations. A natural approach to capsize modeling is to view it as an extremal problem to be handled by the machinery of extreme value theory. However, efforts to do this by fitting specific extreme value distributions, e.g., to maximum roll angle values, have not been overly successful. 
The variety of capsize modes suggests that a variety of modeling approaches may be required. In this section, the so called broaching-to capsize mode will be analyzed using the method proposed in~\citep{lit:leadbetter2}. The goal is to see if the proposed bivariate model can be used for modeling of broaching-to risks.

For a vessel sailing in a following sea, a large overtaking wave may trigger a response which may end in capsizing. There are several ways the capsize event may develop one of these, referred to as broaching-to, results in a sudden change of heading~\citep{lit:spyrou}. In moderate sea states, a vessel is likely to broach-to if it runs with high speed and is slowly overtaken by steep and relatively long waves. However, it may also occur at lower speeds if the waves are steep enough.

In order to assure safe operation of vessels, recommendations are needed for their heading and speed in terms of sea conditions $H_s$ and $T$. These recommendations should be given such that the risk of capsizing is small. 
It is reasonable to assume an exponentially distributed time until capsizing for time scales of hours or larger, since the apparent waves have correlation ranges on much shorter time scales. 
Hence, the risk will be measured by the capsize intensity, which will depend on the type of ship and operating conditions such as sea state, heading, and speed. We summarize the operating conditions in a vector of parameters, $\theta = (H_s, T_z, \alpha, v)$, where $\alpha$ is the angle between the heading of the ship and the direction of the traveling waves, and $v$ is the speed of the ship. 
The angle, $\alpha$, is estimated in the same way as in the fatigue example. 

Let $\lambda(\theta)$ denote the Poisson intensity, meaning the expected number of capsizes in given time unit under the operational conditions $\theta$. 
In order to  estimate capsize probability, a detailed understanding of what constitutes a ``dangerous wave'' is necessary, i.e., what geometrical properties make it more likely to cause capsize when it overtakes a vessel from behind. Further, it seems likely that the probability of such a wave causing a capsize will depend on factors such as the position and motion of the vessel relative to the overtaking wave when an encounter is initiated. 
Simulations on the performance of a Coast guard cutter in severe sea conditions, run by the U.S. Coast Guard, was studied in~\citep{lit:leadbetter2}. For capsizes due to broaching-to, the vessel track of the simulated ship along with the shape of the last wave preceding the capsize event, which we refer to as the ``triggering wave'', were recorded.  A common denominator of the triggering waves is the similar (steep) slope between peak and trough. 
It is therefore reasonable to define a wave as dangerous if its downward slope lies within some range of steep slopes as the wave passes the centre of gravity of the vessel. We then want to calculate the rate $\mu_D(\theta)$ in which dangerous waves are expected to overtake the vessel, and further adjust this by the estimated probability that a dangerous wave will cause a capsize.

\subsubsection{Intensities of potentially dangerous overtaking waves}

A monochromatic plane wave has wavelength $L=2\pi\,g/\omega^2$, period $2\pi \omega^{-1}$, and velocity $V = L/T$. For the ship traveling with speed $v > 0$ and an angle of $-\pi/2 < \alpha< \pi/2$ to the propagating direction of the wave, the intensity of overtaking waves is $\mu(\theta)=(V-v_x)^+/L$.
Here, $v_x = v\cos(\alpha)$ and $a^+ = \max(a,0)$. Note that a wider angle between the heading of the vessel and the wave direction yields a higher intensity. Likewise, a smaller ship speed also yields a higher intensity. At the same time, too large values of $\alpha$ will not cause dangerous broaching-to events since the heading of the ship will not change dramatically; although encountering big waves perpendicular to the heading of a ship can be dangerous for other reasons. 

Similarly to the monochromatic wave, the intensity of an apparent wave overtaking the center of gravity of the ship in a non-degenerate Gaussian sea has been shown to be~\citep{lit:rychlik3}  
\begin{equation}
    \mu(\theta)=\expect{\frac{(V-v_x)^+}{L}}=\frac{1}{4\pi}\sqrt{{\frac{m_{20}}{m_{00}}}}
    \left( -\frac{m_{11}}{m_{20}} - v_x + \sqrt{v_x^2+2\frac{v_xm_{11}}{m_{20}}+\frac{m_{02}}{m_{20}} } \right),
    \label{eq:mu}
\end{equation}
where 
\begin{align}
    m_{ij} := \int_0^{\infty} \left( \frac{\omega^2}{g} \right)^i \omega^j S(\omega)d\omega 
\end{align}
are the spectral moments of the Gaussian process.

A ship being overtaken by an apparent wave is only dangerous if the wave is high and has a steep slope. 
Analytic derivations~\citep[Theorem 6.2]{lit:aberg} give an explicit formula for the CDF of $W_x(x_0, t_0)$, where, $x_0, t_0$ are instances in space-time where the center of gravity of the ship is being overtaken by the zero level down-crossing of an apparent wave, and $W_x$ is the partial derivative of $W$ with respect to the spatial direction of the propagating wave. 
The formula for the CDF is
\begin{equation}
F_{W_x}(r)=
\begin{cases}
\frac{2}{1-\rho}\left(\Phi(r/\sigma)-\rho\,e^{-\frac{r^2}{2m_{20}}}\Phi(r\rho/\sigma)\right)&,\quad r \le 0 \\
1&, \quad r > 0.
\end{cases}
\label{eq:slope}
\end{equation}
Here, $\Phi(x)$ is the CDF of the standard normal distribution, $\sigma^2=m_{20}(1-\rho^2)$, and
\begin{align}
\rho=\frac{v_xm_{20}+m_{11}}{\sqrt{m_{20}(v_x^2m_{20}+2v_xm_{11}+m_{02})}}.
\end{align}

The intensity of a broaching-prone wave scenario is the product of the intensity of overtaking waves thinned with the probability that the overtaking wave has a dangerously steep slope, i.e.,
\begin{equation}
\mu_D(\theta)=\mu(\theta)\prob{ W_x(x_0, t_0) \in A },
\label{eq:muD}
\end{equation} 
where $A$ is an interval of slopes considered dangerous.
Inspired by~\citet{lit:leadbetter2}, we choose $A = [-0.4, -0.2]$. 

Since the spectral moments are known functions of $H_s$ and $T$, assuming a Bretschneider spectrum, we can compute them for each point on the route for a given realization of $H_s$ and $T$. 
In the following example we computed the spectral moments assuming a limited bandwidth and numerical integration using the Matlab toolbox WAFO~\citep{lit:brodtkorb}. 
Using the route of Figure~\ref{fig:route} and wave directions of Figure \ref{fig:waveDirection}, $\mu_D(\psp_{\gamma}(t))$ can be estimated conditioned on a given sea state scenario.

\subsubsection{Estimation of $\lambda(\theta)$ response surfaces.}

Conditioned on the ship being overtaken by a ``dangerous'' wave, the capsizing phenomenon is a result of complicated nonlinear interactions between the wave and the vessel. Direct computations of risk for capsizes based on random models for sea motion and vessel response are not feasible to obtain. 
In addition, there are limited data of capsizing available. 
Consequently, one must study the problem using tank experiments with model ships or by means of computer simulations of the responses.
Since a capsize due to broaching-to occurs with a small probability, tank experiments would require too much time to get stable estimates of capsize probability for all but the most severe sea states. Instead, appropriate computer simulations are the best methods for estimating the probability of capsize and related events under moderately high sea conditions. 

\citet{lit:leadbetter2}~derived a method for modeling the capsizing intensity due to broaching-to, $\lambda$, based on Poisson regression on the covariates $\mu_D$, $H_s$, and $T_1$, i.e.,
\begin{align}
    \lambda(\theta) = \mu_D(\theta) \exp\left(\beta_0 + \beta_H \log H_s + \beta_T \log T\right).
\end{align}
The values of $\beta_0, \beta_H$, and $\beta_T$ depend on the ship type in consideration; a heavier and larger ship can withstand taller waves without broaching-to, as compared to a small ship.
The parameters of the regression for a U.S. coast guard cutter were estimated in~\citep{lit:leadbetter2}. 
It turned out that this standard linear Poisson regression satisfactorily explained $\lambda(\theta)$ with the parameters $\beta_0, \beta_H,$ and $\beta_T$ estimated from capsize data in the computer simulations. 
The values were $\beta_0 \approx \log(0.05), \beta_H \approx 7.5,$ and $\beta_T \approx -7.5$.
The model was shown to predict intensities of order $10^{-3}$ adequately. 
It is still not known if the model can be extrapolated to even safer operating conditions. However, the predicted sea states that should be avoided are in line with the ones found using significant roll threshold, see~\citep[Fig. 22.2]{lit:leadbetter2}.

For a ship traversing the route of Figure~\ref{fig:route}, $\lambda(\theta(\psp_{\gamma}(t)))$ is the conditional capsize intensity of an inhomogeneous Poisson process over the space-time curve of the ships path, given the sea states, $\theta$.
The distribution of capsizes, if assuming that a ship could continue after a capsize, would then be Poisson distributed with intensity,
\begin{align}
    \lambda(\theta) := \int_{0}^{t_{end}}\lambda(\theta_{\gamma}(t)) dt \approx \sum_{i=1}^{100}\lambda(\theta_{\gamma}(t_i))\Delta t,
\end{align}
where $\theta_{\gamma}(t) := \theta(\psp_{\gamma}(t))$.
The capsize events can hence be considered as a Cox process where the latent random intensity is given by the sea states, $\theta$.

In our example we compute the distribution of $\lambda$ as a function of the bivariate random field $H_s, T$.  We use the same coefficients as in~\citet{lit:leadbetter2}, i.e., $\beta_0 = \log(0.05), \beta_1 = 7.5, \beta_2 = -7.5$. 
When computing $\lambda$ we consider traversing the route from America to Europe, with the wave directions as in Figure \ref{fig:waveDirection}. 
Furthermore, we choose the cutoff angle, $\alpha_0 = 75^{\circ}$, meaning that we only consider waves as potentially dangerous if the angle between the ships heading and the propagation direction of the waves are less than $\alpha_0$.
The scenario of traversing the route from Europe to America was not considered since the wave direction angle was always more than $\alpha_0$, i.e., negligible risk of a dangerous apparent wave overtaking the ship from behind. 

\begin{figure}[tp]
    \centering
    
    \begin{subfigure}{0.45\textwidth}
        \centering
        Bivariate spatial model
    \end{subfigure}
    \begin{subfigure}{0.45\textwidth}
        \centering
        Univariate spatial model
    \end{subfigure}    
    \vspace{ 0.5 cm}
    
    \begin{subfigure}{0.45\textwidth}
        \centering
        \includegraphics[width = \textwidth]{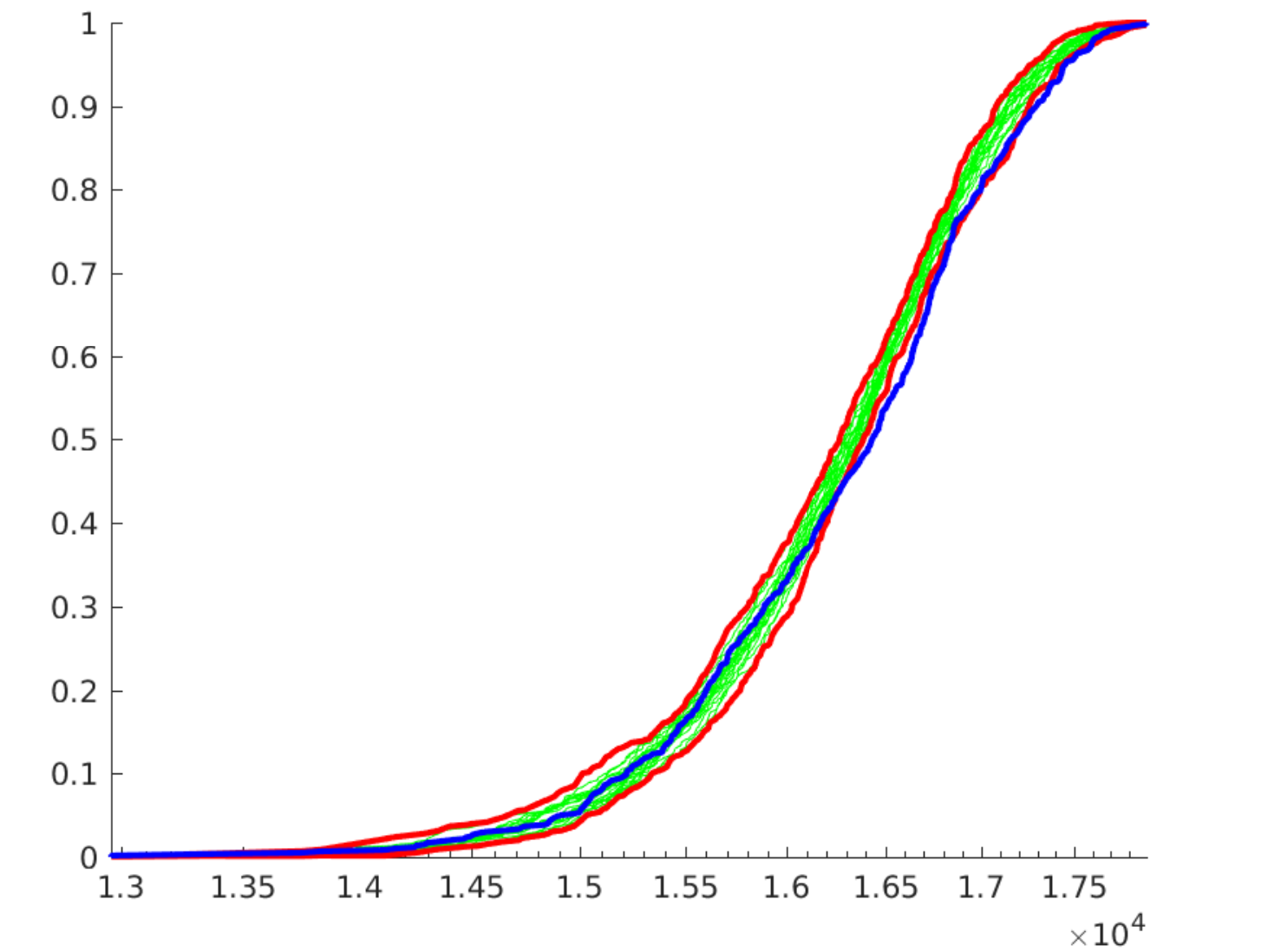}
        \caption{$\mu$ bivariate}
    \end{subfigure}
    \begin{subfigure}{0.45\textwidth}
        \centering
        \includegraphics[width = \textwidth]{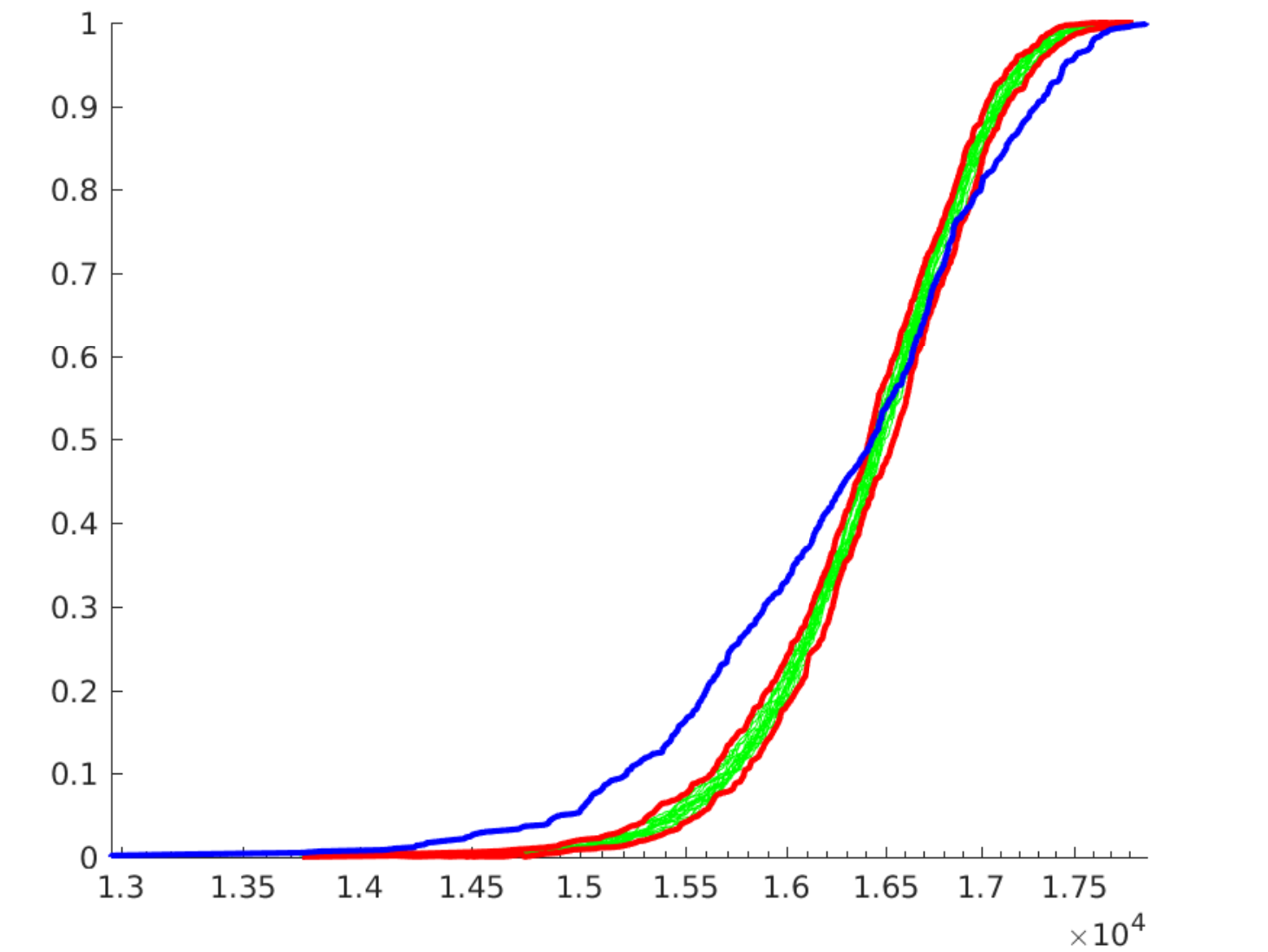}
        \caption{$\mu$ univariate}
    \end{subfigure}
    
    \begin{subfigure}{0.45\textwidth}
        \centering
        \includegraphics[width = \textwidth]{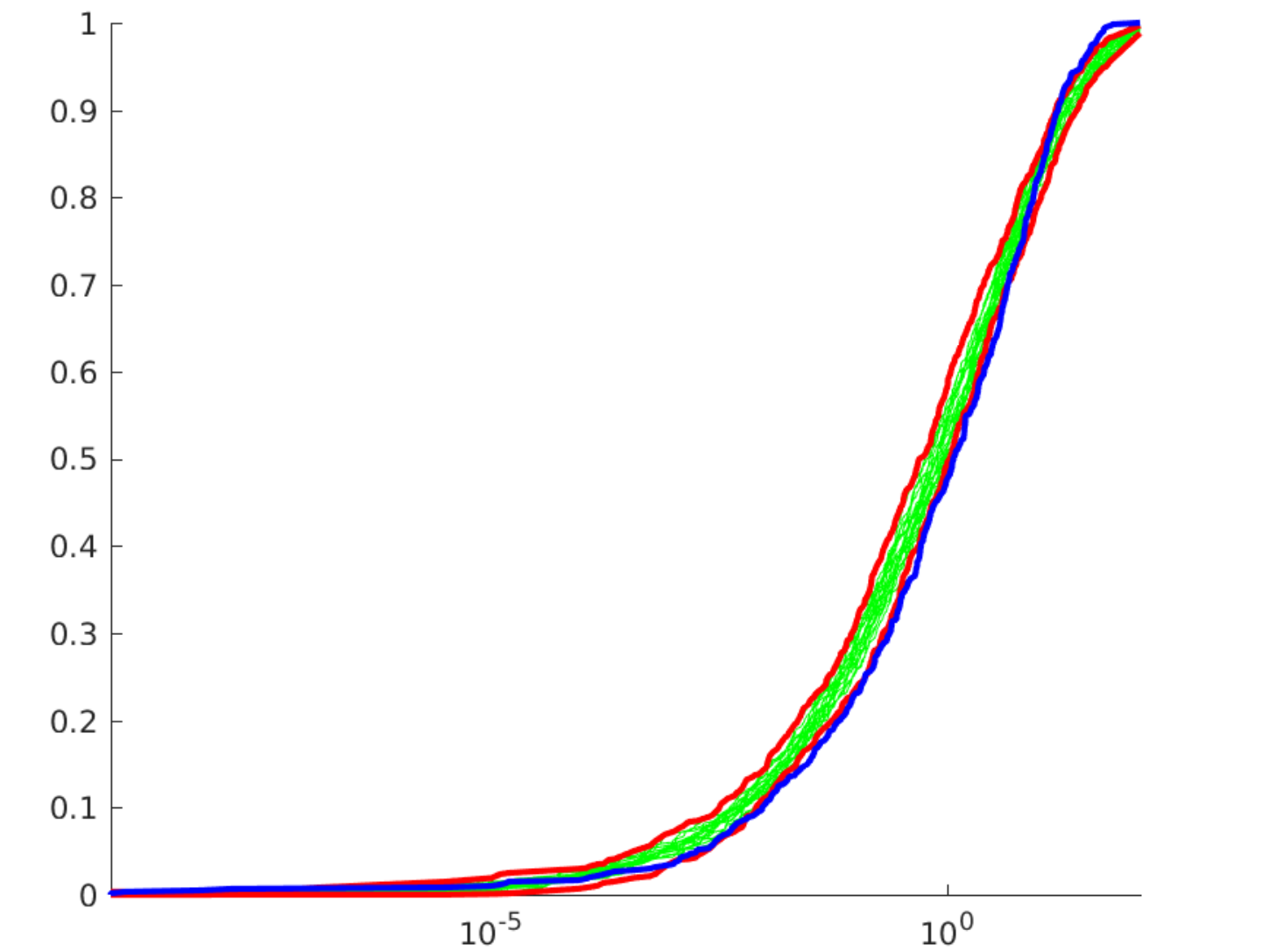}
        \caption{$\mu_D$ bivariate}
    \end{subfigure}
    \begin{subfigure}{0.45\textwidth}
        \centering
        \includegraphics[width = \textwidth]{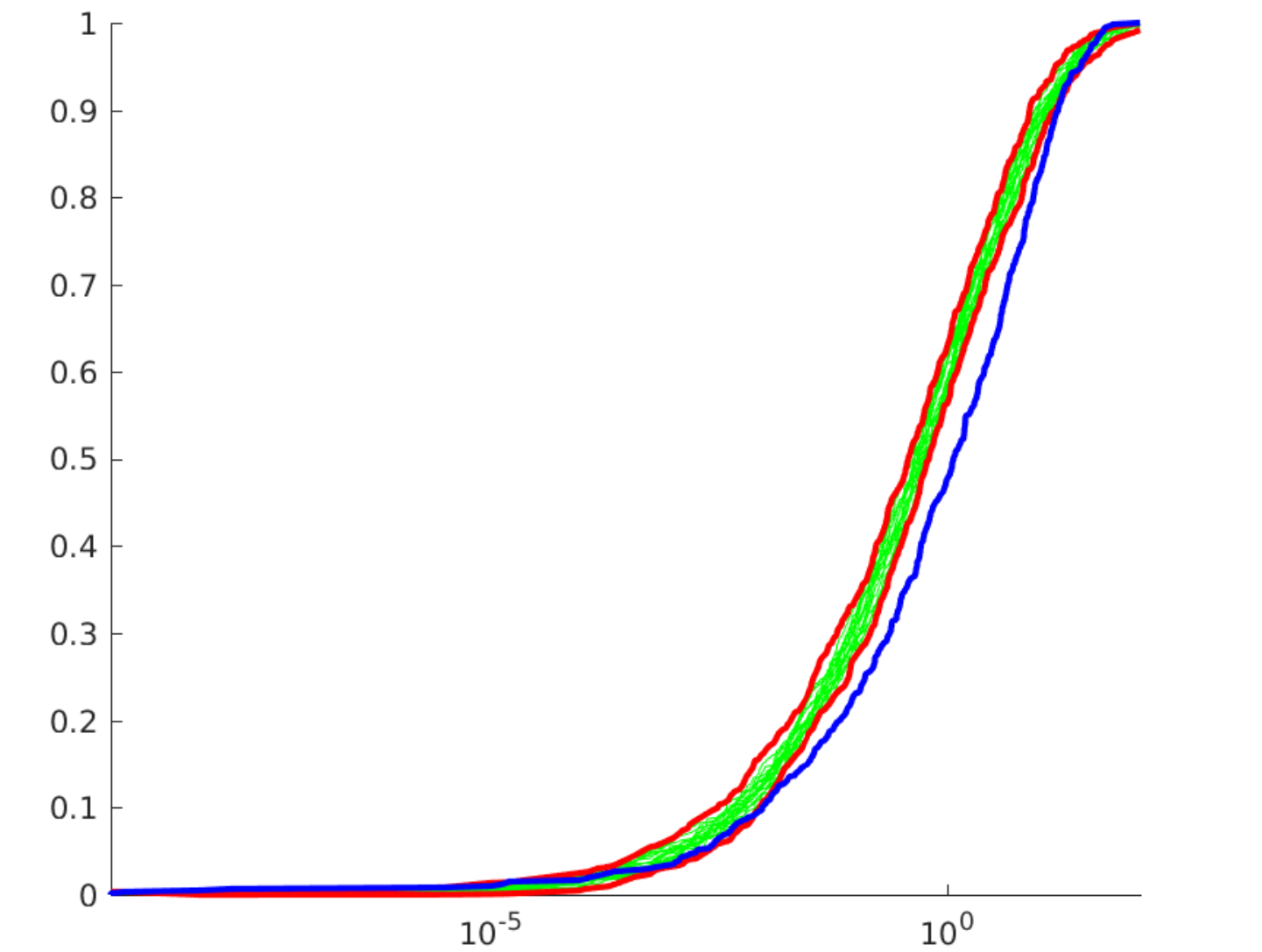}
        \caption{$\mu_D$ univariate}
    \end{subfigure}
    
    \begin{subfigure}{0.45\textwidth}
        \centering
        \includegraphics[width = \textwidth]{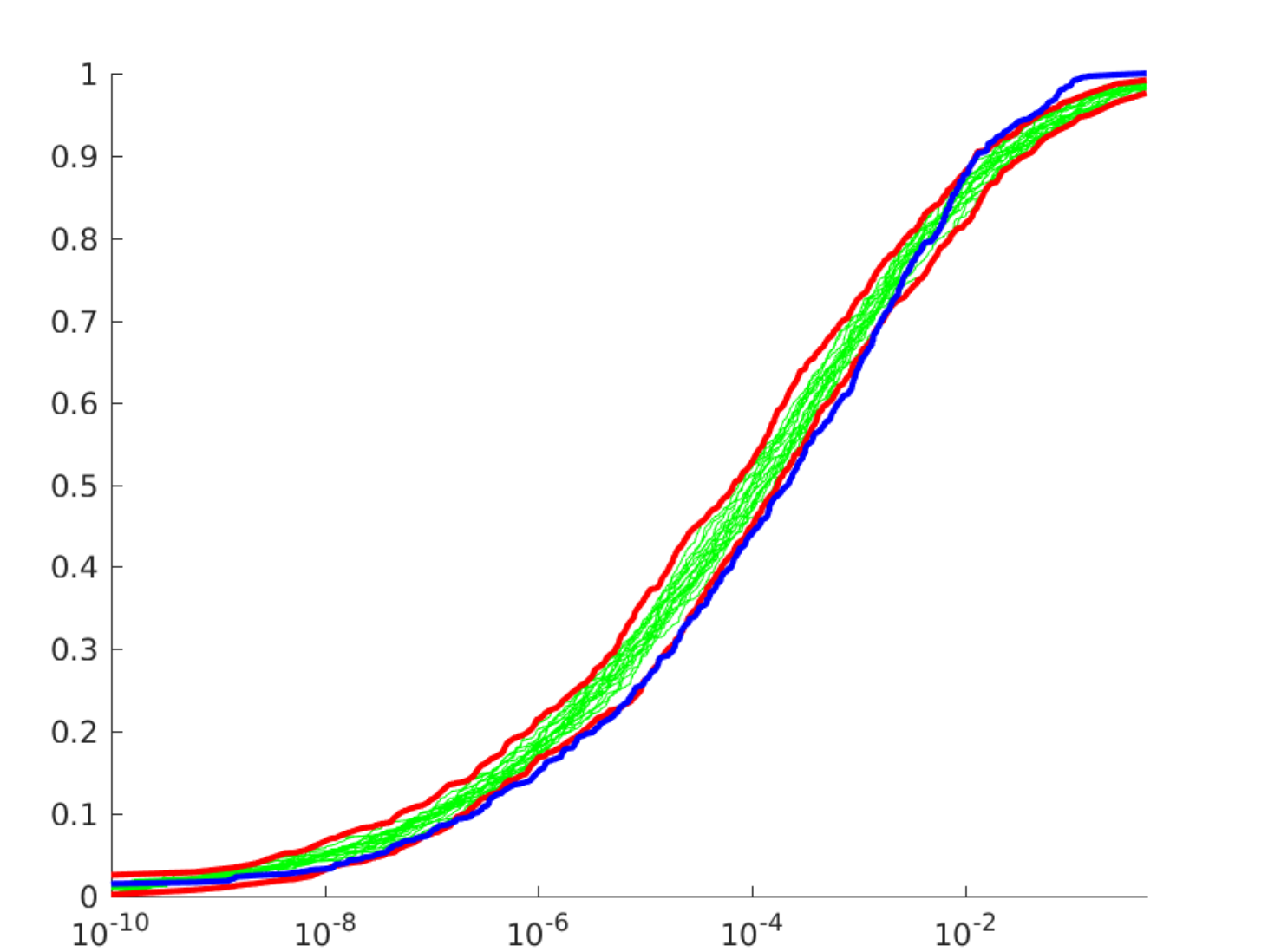}
        \caption{$\lambda$ bivariate}
    \end{subfigure}
    \begin{subfigure}{0.45\textwidth}
        \centering
        \includegraphics[width = \textwidth]{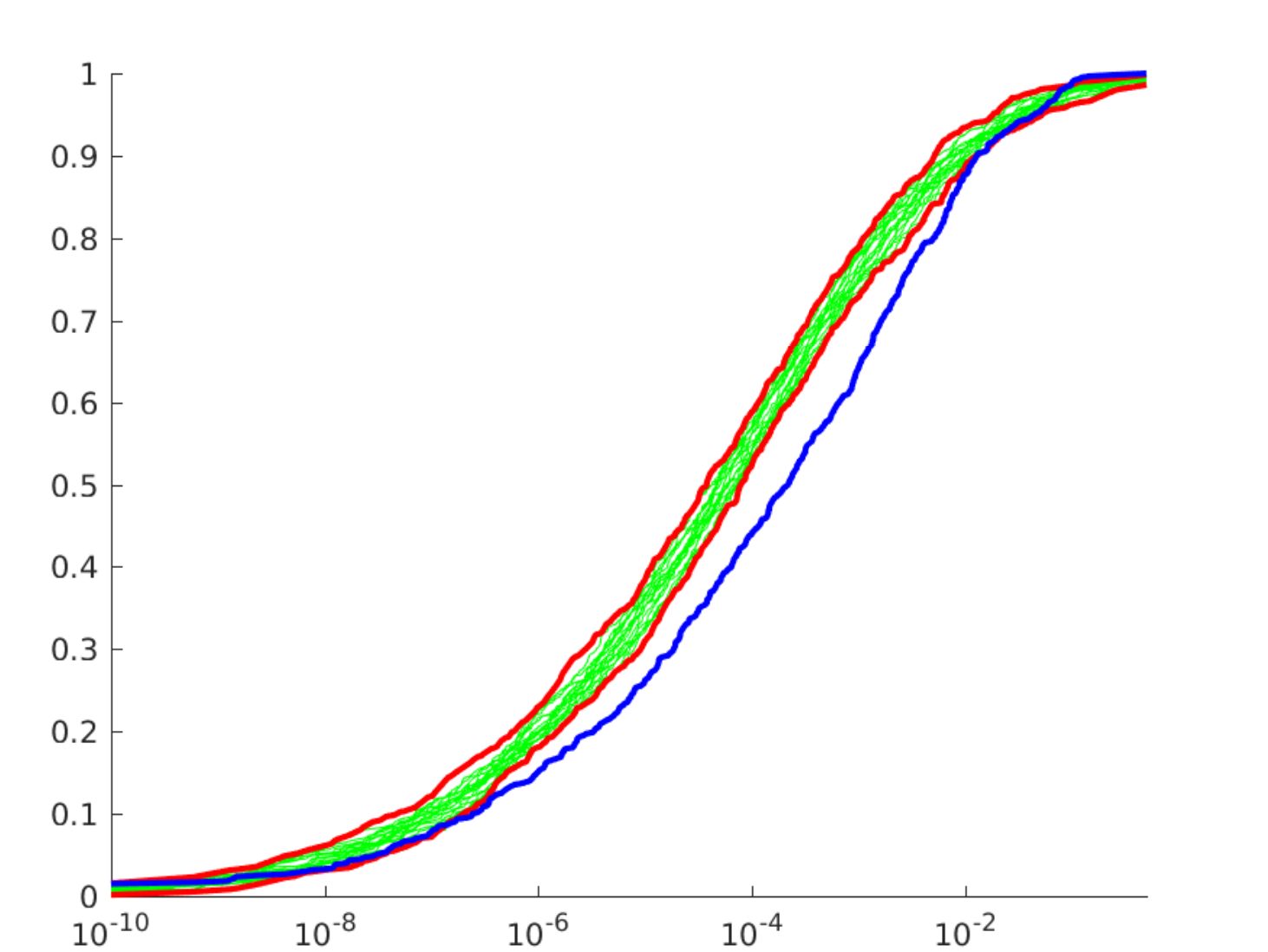}
        \caption{$\lambda$ univariate}
    \end{subfigure}
    
    \caption{Empirical CDFs of total $\mu$, $\mu_D$, and $\lambda$. The broaching-to risks were computed using two different models of sea states, the proposed bivariate random field model (left) and the simpler model of spatial $H_s$ with the marginal conditional mean of $T$ (right). Empirical CDF from data (blue), corresponding empirical CDFs from 20 simulations (green) and the pointwise upper and lower envelopes of the empirical CDFs (red). }
    \label{fig:broaching}
\end{figure}

The distribution of capsize intensities, $\lambda$, as well as corresponding total intensities of overtaking waves and dangerous overtaking waves can be seen in Figure~\ref{fig:broaching}.
The figure shows the estimated CDF of the total intensities, $\mu$, $\mu_D$, and $\lambda$, for a ship traversing the transatlantic route of Figure~\ref{fig:route} from America to Europe. The left column correspond to computations using the proposed bivariate spatial random model of sea states. The right column correspond to the simpler model of the univariate spatial $H_s$ model together with the pointwise conditional mean of $T$, which was found to be sufficient for the fatigue application in Section~\ref{sec:fatigue}. The CDF computed from the data is compared with 20 simulations of equal size, 600 days. 

As is seen in Figure~\ref{fig:broaching}, the simpler model is now clearly deviating from the empirical CDF of the data. The proposed bivariate spatial model show a better fit although it seems to overestimate the risks slightly for medium sized intensities. Thus, for this application the bivariate model is clearly outperforming the simpler alternative.

\section{Discussion}
\label{sec:discussion}

A joint spatial model of significant wave height and wave period has been introduced. The model is a bivariate extension of the model of~\citet{lit:hildeman} using the multivariate random field approach of~\citet{lit:bolin, lit:hu}. Furthermore, the model also incorporates the rational approximation to Mat\'ern fields of arbitrary smoothness~\citep{lit:bolin3}. 
This means that the spatial model allows for non-stationary, anisotropic models of bivariate Gaussian random fields, each with its own arbitrary smoothness. The model is parametrized with a relatively small number of easily interpretable parameters.   

The model was fitted using data from the month of April from the ERA-Interim global atmospheric reanalysis \citep{lit:dee}. 
A stepwise maximum likelihood approach together with numerical optimization by a quasi-Newton method was used to estimate the parameters of the model. 
The univariate models for $H_s$ and $T$ separately agrees well with data. 
However, problems were encountered when fitting the cross-correlation structure between $H_s$ and $T$.
The problem is that the cross-correlation is not at its maximum between the same spatial points in $H_s$ and $T$, as assumed by the model. This lead to ML~estimates of the cross-correlation structure that did not agree at all with the observed data. Instead, estimating the cross-correlation structure using a pointwise maximum likelihood method yielded better results; although the cross-correlation range was clearly underestimated for small values. 

The shift of locations of maximum cross-correlation, as seen in Figure~\ref{fig:crosscorrelationTranslation}, is likely an effect of the dynamic nature of ocean waves and their interaction with wind. 
The proposed model assumes a symmetric cross-correlation structure with maximum cross-correlation between the same point in the two fields. 
Due to the shifts, the real cross-correlation is not symmetric. 
Because of this, it would make sense to incorporate these shifts into the bivariate model using the model of~\citet{lit:li}. This is an interesting extension of the multivariate modeling approach using systems of SPDEs and was proposed in~\citep{lit:hu2}.
Such shifts could be considered as a diffeomorphism between $\gspace$ and some overlapping region $\mathcal{H}$. 
This diffeomorphism would fulfill that when $\log T$ is mapped to $\mathcal{H}$, the two fields, $\log H_s$ and $\log T$, align, i.e., maximum cross-correlation is between the same point in the two fields. The proposed model of this paper could then be applied to this transformed data.

The spatial model was evaluated in two applications in naval logistics. Both applications considered risks of undertaking a journey between the European and American continents through the north Atlantic. The first application considered computing the probability distribution of accumulated fatigue damage acquired during the journey. It was shown that the spatial model agreed with data. In particular, it showed that it works better than the approach where $T_z$ is replaced by the proxy $T_z = 3.75\sqrt{H_s}$, which was used in~\citep{lit:hildeman}.
However, a simpler model using only the univariate spatial random field model of $H_s$ together with pointwise conditional means of $T$ given $H_s$ yielded an adequate fit as well. 

The second application concerned the risk of capsizing due to broaching-to. An inhomogeneous Poisson process was derived given the bivariate sea state surface of $H_s$ and $T$. The Poisson intensity depended on the intensity of the ship being overtaken by a wave from behind, the probability that the overtaking wave is steep, and a Poisson regression of the probability of capsizing given a dangerous wave.
The distribution of capsizing intensity (corresponding to the risk of capsizing) was compared between the proposed bivariate spatial model and the data. The spatial model showed a reasonable fit but seems to overestimate the risk slightly. 
The simpler model, using the univariate random field of $H_s$ from~\citep{lit:hildeman} together with the pointwise conditional mean of $T$, was on the other hand clearly deviating from the distribution of the data. This shows that the bivariate model is indeed important for certain applications, and cannot simply be substituted by simpler univariate models.

\section{Acknowledgements}
We would like to thank the European Centre for Medium-range Weather Forecast (ECMWF) for the development of the ERA-Interim data set and for making it publicly available. 
The data used was the ERA-Interim reanalysis dataset, Copernicus Climate Change Service (C3S) (accessed September 2018), available from ``https://www.ecmwf.int/en/forecasts/datasets/archive-datasets/reanalysis-datasets/era-interim''.

\bibliography{ms}

\end{document}